\documentclass[twocolumn]{aastex6}

\usepackage{natbib}
\usepackage[caption=false]{subfig}
\usepackage{amsmath,amssymb}

\newcommand{\degree}{$^\circ$}

\newcommand{\kms}{km~s$^{-1}$}

\newcommand{\rjup}{R$_{Jup}$}

\newcommand{\lbol}{$\log_{10}{L_{bol}/L_{\sun}}$}
\newcommand{\teff}{T$_{eff}$}
\newcommand{\logg}{$\log{g}$}

\newcommand{\sha}{2MASS~J0835$-$0819}
\newcommand{\shb}{2MASS~J1821+1414}

\slugcomment{To be submitted to ApJ/AJ}

\shorttitle{Spectral Variability of L Dwarfs}
\shortauthors{Schlawin et al.}

\begin{document}

\title{Spectral Variability of Two Rapidly Rotating Brown Dwarfs: \\2MASS~J08354256$-$0819237 and 2MASS~J18212815+1414010}


\author{E. Schlawin\altaffilmark{1,2}, A. J.\ Burgasser\altaffilmark{2,3}, T. Karalidi\altaffilmark{1}, J.\ E.\  Gizis\altaffilmark{4},  \and J. Teske\altaffilmark{5}}

\altaffiltext{1}{Steward Observatory, Tucson AZ 85721 \email{eas342@email.arizona.edu}}
\altaffiltext{2}{Visiting Astronomer at the Infrared Telescope Facility, which is operated by the University of Hawaii under contract NNH14CK55B with the National Aeronautics and Space Administration}
\altaffiltext{3}{Center for Astrophysics and Space Science, University of California San Diego, La Jolla, CA 92093, USA}
\altaffiltext{4}{Department of Physics and Astronomy, University of Delaware, Newark, DE 19716, USA}
\altaffiltext{5}{Carnegie Observatories, 813 Santa Barbara Street, Pasadena, CA 91101, USA}

\begin{abstract}
L dwarfs exhibit low-level, rotationally-modulated photometric variability generally associated with heterogeneous, cloud-covered atmospheres. The spectral character of these variations yields insight into the particle sizes and vertical structure of the clouds. Here we present the results of a high precision, ground-based, near-infrared, spectral monitoring study of two mid-type L dwarfs that have variability reported in the literature, 2MASS~J08354256$-$0819237 and 2MASS~J18212815+1414010, using the SpeX instrument on the Infrared Telescope Facility. By simultaneously observing a nearby reference star, we achieve $<$0.15\% per-band sensitivity in relative brightness changes across the 0.9--2.4$\mu$m bandwidth. We find that 2MASS~J0835$-$0819 exhibits marginal ($\lesssim$ 0.5\% per band) variability with no clear spectral dependence, while 2MASS~J1821+1414 varies by up to $\pm$1.5\% at 0.9~$\mu$m, with the variability amplitude declining toward longer wavelengths. The latter result extends the variability trend observed in prior $HST$/WFC3 spectral monitoring of 2MASS~J1821+1414, and we show that the full 0.9--2.4~$\mu$m variability amplitude spectrum can be reproduced by Mie extinction from dust particles with a log-normal particle size distribution with a median radius of 0.24 $\mu$m. We do not detect statistically significant phase variations with wavelength. The different variability behavior of 2MASS~J0835$-$0819 and  2MASS~J1821+1414 suggests dependencies on viewing angle and/or overall cloud content, underlying factors that can be examined through a broader survey.
\end{abstract}

\keywords{techniques: spectroscopic -- stars: atmospheres -- stars: brown dwarfs -- stars: individual (\objectname{2MASS J08354256-0819237}, \objectname{2MASS J18212815+1414010}) -- stars: late-type -- stars: variables: general}

\section{Introduction}

The atmospheres of L-type stars and brown dwarfs (1300~K $\lesssim$ {\teff} $\lesssim$ 2200~K; \citealt{2015ApJ...810..158F}) are sufficiently cool that species of mineral and metal condensates form spontaneously in their atmospheres \citep{1996A&A...305L...1T,2010ApJ...716.1060V}.
In addition to influencing spectral energy distributions (e.g., \citealt{2001ApJ...556..357A,2008ApJ...674..451B}) and photospheric chemistry (e.g., \citealt{2000ApJ...531..438B}), condensates can coalesce into large-scale cloud and haze structures \citep{1989ApJ...338..314L,2001ApJ...556..872A,2014Natur.505..654C}.
These structures may produce rotationally-modulated brightness variations (e.g., \citealt{2012ApJ...750..105R}) and long-term variations from differential rotation and/or cloud structure evolution (e.g., \citealt{2009ApJ...701.1534A,2013A&A...555L...5G,2016ApJ...826....8Y}). Evidence of cloud-driven variability extends to the cooler T dwarfs (e.g., \citealt{2009ApJ...701.1534A,2012ApJ...760L..31B,2015ApJ...799..154M}); virtually all L dwarfs and most T dwarfs likely have cloud deck heterogeneities leading to infrared flux variations $\gtrsim$0.2\% in amplitude \citep{2015ApJ...799..154M}. Clouds and hazes are important constituents of exoplanetary atmospheres as well (e.g., \citealt{2011ApJ...733...65B,2014Natur.505...69K,2016Natur.529...59S}).

Multi-wavelength monitoring of low-temperature stars and brown dwarfs allows for the measurement of cloud vertical structure and composition, as gray grain scattering opacity competes with strongly wavelength-dependent molecular gas opacity. The amplitude of variability as a function of wavelength samples the vertical depths of cloud layers \citep{2013ApJ...768..121A}, and amplitude spectra have been shown to vary with effective temperature ({\teff}; e.g., \citealt{2015ApJ...798L..13Y}) and surface gravity (e.g., \citealt{2016ApJ...829L..32L}).
Wavelength-dependent phase variations in lightcurves have also been observed \citep{2012ApJ...760L..31B,2013ApJ...778L..10B,2016ApJ...826....8Y}.
Since different wavelengths sample different photospheric depths in the highly structured spectra of L and T dwarfs, these phase variations have been interpreted as arising from the relative motion of distinct cloud layers or deep thermal pulsations. Deep thermal pulsations occur on longer timescales ($\sim$ 100 hr for an example T-dwarf model) but may interact with clouds to produce shorter timescale variability \citep{2014ApJ...785..158R}.

Broad-band variability amplitudes are typically $<$1\% in L and T dwarfs, with some dramatic exceptions \citep{2009ApJ...701.1534A,2012ApJ...750..105R,2013A&A...555L...5G,2016ApJ...829L..32L}. 
Ground-based spectral (e.g., CLOUDS; \citealt{2008A&A...487..277G}) and broad-band photometric monitoring surveys (e.g., \citealt{1999A&A...348..800B,2003MNRAS.346..473K,2014ApJ...793...75R,2014A&A...566A.111W}) are generally limited to $>$1-3\% relative precision due to telluric atmospheric instabilities (i.e., ``red noise'') and insufficient numbers of nearby reference stars which can be used to correct time-variable telluric absorption \citep{2003MNRAS.339..477B}.
In contrast, the stability of space-based platforms such as $Hubble$ $Space$ $Telescope$ ($HST$), $Spitzer$ and $Kepler$ provide order-of-magnitude improvements in relative photometric and spectro-photometric precision and have been responsible for the discovery of the majority of known L and T dwarf variables \citep{2013ApJ...768..121A,2013ApJ...779..172G,2015ApJ...799..154M}. However, these facilities are poorly suited for building up robust statistical samples to assess the physical origins of surface heterogeneity or evolution in variability modes. 

Ground-based Multi-Object Spectroscopy (MOS) with a multi-object or long slit spectrograph, in which reference stars are simultaneously observed to correct for telluric effects can significantly improve the precision of ground-based spectro-photometry (e.g., \citealt{bean10,bean2013,gibson13clouds,stevenson2016hatp26}). \citet{2014ApJ...783....5S} achieved $<$0.1\% per band variability precision for the planet-host star CoRoT-1 using the NASA Infrared Telescope Facility (IRTF) SpeX spectrograph \citep{2003PASP..115..362R}, by simultaneously monitoring the star and a nearby reference star in a wide and long (3$\arcsec \times 60 \arcsec$) slit.
This was sufficient to infer the presence of a cloud/haze layer or disfavor TiO/VO absorption in CoRoT-1b via transmission spectroscopy. A similar investigation of the disintegrating rocky exoplanet KIC 12557548b  enabled the characterization of the grain properties of its tidal debris tail \citep{2016ApJ...826..156S}. In a parallel effort, \citet{2014ApJ...785...48B} obtained $<$0.5\% per-band relative precision across the 1--2.5~$\mu$m band with IRTF/SpeX for the variable T dwarf Luhman 16B, by simultaneously observing Luhman 16A, a physically associated L dwarf companion nearby on the sky. These studies demonstrate that high-precision, near-infrared spectral monitoring observations can be achieved from the ground.

\begin{deluxetable*}{llllllllllr}
\tablecaption{Brown Dwarf Properties}\label{tab:bdProp}
\tablewidth{0pt}
\tabletypesize{\scriptsize}
\tablehead{
\colhead{Name} & 
\multicolumn{2}{c}{Spectral Type} & 
\colhead{$d$} & 
\colhead{2MASS $H$} & 
\colhead{$J-K_s$} & 
\colhead{\lbol} & 
\colhead{\teff} & 
\colhead{P$_{rot}$} &
\colhead{$v\sin{i}$} & 
\colhead{$i$\tablenotemark{b}} \\
 &  \colhead{Opt} & \colhead{NIR\tablenotemark{a}} &  \colhead{(pc)} & &  & &  \colhead{(K)} & \colhead{(hr)} & \colhead{(km/s)} & \colhead{($\degr$)} \\
 }
\startdata
\sha\ & L5$^{(1)}$ & L7 & 7.27$\pm$0.02$^{(2)}$ & 11.94$\pm$0.02 & 2.03$\pm$0.03 & $-$4.05$\pm$0.08$^{(3)}$ & 1754$\pm$112$^{(3)}$ & 3.1\tablenotemark{c}$^{(4)}$ & 14.18$\pm$0.43$^{(5)}$ & 21 \\
\shb\ & L4.5p$^{(6)}$ & L5 & 9.38$\pm$0.03$^{(7)}$ & 12.40$\pm$0.02 & 1.78$\pm$0.03 & $-$4.18$\pm$0.03$^{(8)}$ & 1635$\pm$66$^{(8)}$ & 4.2$\pm$0.1$^{(9)}$ & 28.85$\pm$0.16$^{(5)}$ & 76 \\
\enddata
\tablenotetext{a}{This Work.}
\tablenotetext{b}{Estimated inclination angle $i$ assuming a radius = 1 Jupiter radius, so that $\sin{i}$ = 8.0$\times$10$^{-3}Pv\sin{i}$, where $P$ is in hours and $v\sin{i}$ is in km/s. A value of $i$ $\approx$ 0{\degree} is viewed pole-on, $i$ $\approx$ 90{\degree} is viewed equator-on.}
\tablenotetext{c}{Tentative period.}
\tablerefs{(1) \citet{2003AJ....126.2421C};
(2) \citet{2016AJ....152...24W}; 
(3) \citet{2015ApJ...810..158F};
(4) \citet{2004MNRAS.354..378K};
(5) \citet{2010ApJ...723..684B};
(6) \citet{2008ApJ...686..528L}; 
(7) \citet{2016MNRAS.455..357S};
(8) \cite{2016ApJS..225...10F} with updated values as discussed in Section \ref{sec:targets};
(9) \citet{2015ApJ...799..154M}
}
\end{deluxetable*}

\begin{figure*}
\centering
	\plottwo{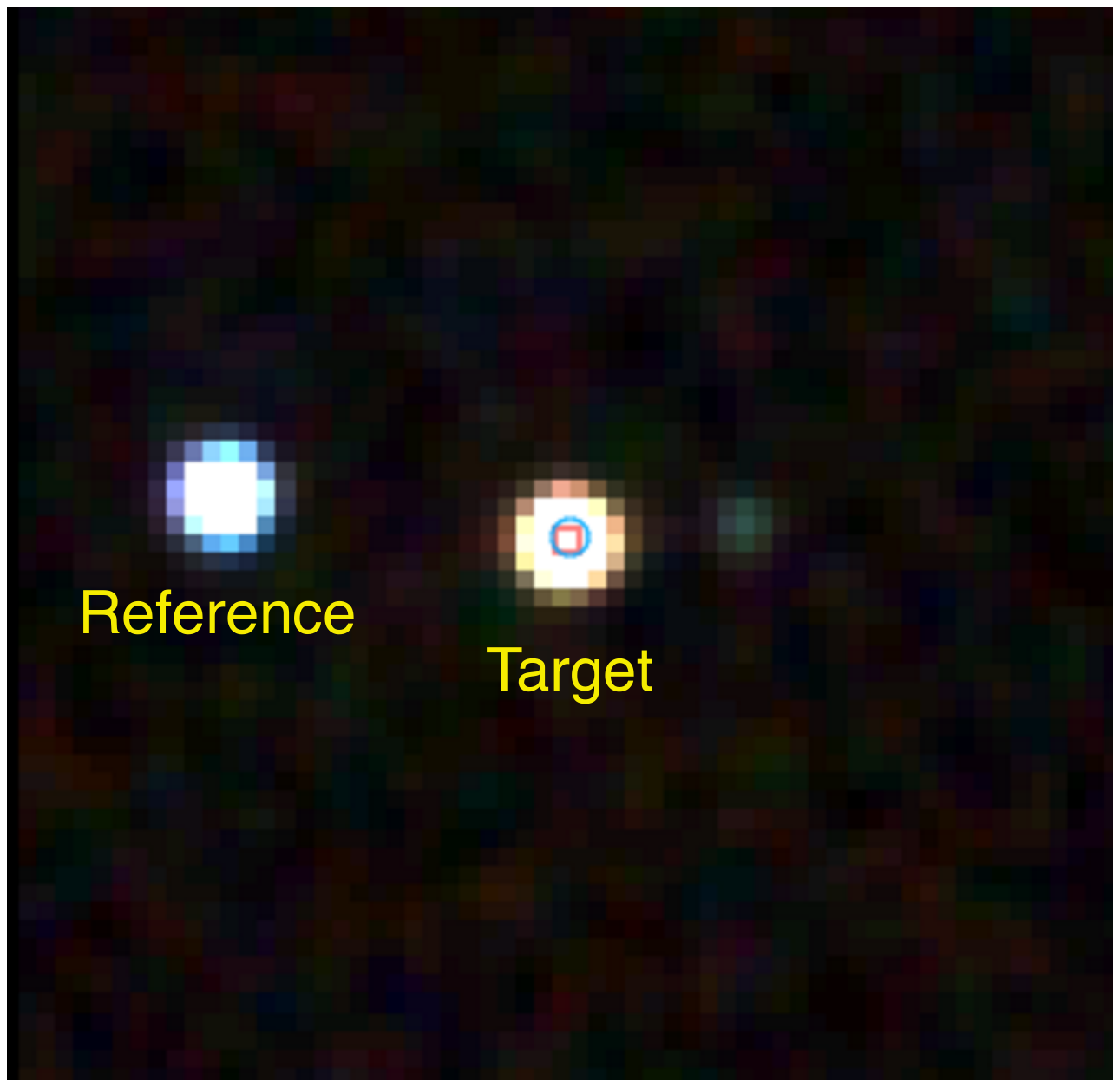}{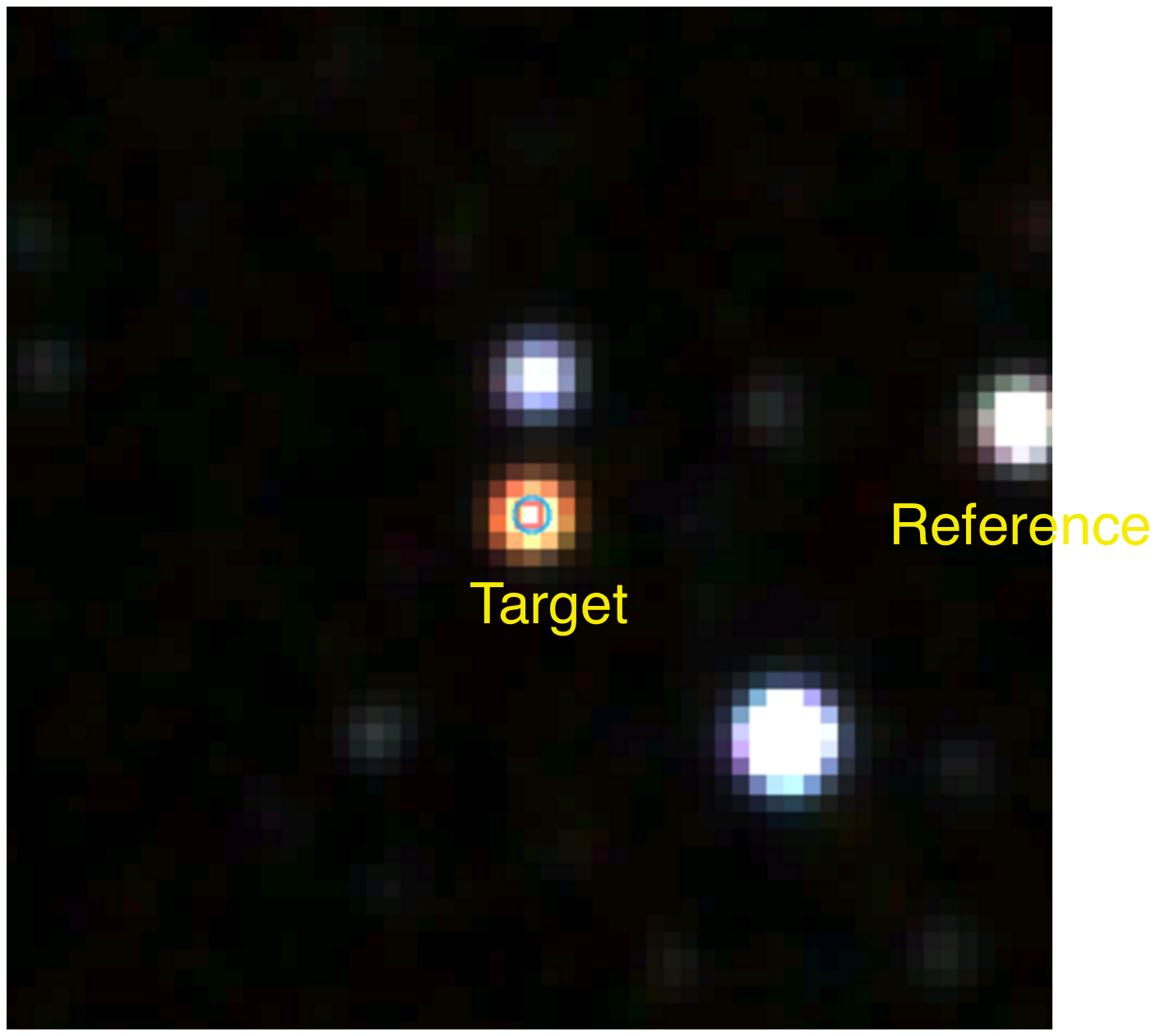}
	\label{fig:image1821}
	\vspace{-.9in}
	\caption{2MASS $JHK_s$ three-color images of the fields around our targets {\sha} (left) and {\shb} (right)
	from the IRSA Finderchart service (\url{http://irsa.ipac.caltech.edu/applications/finderchart/}).  Each image is centered on the target and spans a 1$\arcmin\times$1$\arcmin$ field of view oriented with North up and East the left.
	The reference stars labeled in each image are observed simultaneously in the 3$\arcsec \times 60 \arcsec$ SpeX slit to remove telluric variability.}
	\label{fig:images}
	\vspace{0.17in}
\end{figure*} 

This article reports the results of a pilot study adapting these methods to measure the variability of two L-type dwarfs with nearby visual companions. We describe our target selection process and pilot target properties in Section~\ref{sec:targets}.
In Section~\ref{sec:obs}, we describe our monitoring observations and data reduction procedures to generate relative spectro-photometric time series.
In Section~\ref{sec:analysis}, we analyze these time series using sinusoidal model fits over discrete spectral bands to generate individual light curves and amplitude spectra.
We then fit the variability amplitude spectra with a Mie extinction model for the clouds and constrain the particle size distribution.
We summarize our results in Section~\ref{sec:conclusions}.

\section{Target Selection and Characterization}\label{sec:targets}

We identified potential targets for relative spectro-photometry by selecting among the $\approx$200 known L dwarfs\footnote{We identified these sources from DwarfArchives,  \url{http://dwarfarchives.org}. The declination range corresponds to the visibility region of IRTF.} with $H$ $<$ 14 and $-$30{\degree}  $<$ $\delta$ $<$ +67{\degree}.
We down-selected those sources that have comparable-brightness ($\Delta{H}$ $<$ 1) reference stars at apparent separations of 5$\arcsec$--30$\arcsec$.
We identified 30 potential targets and prioritized those with previously detected photometric variability or rotational $v\sin{i}$ measurements indicating periods $\leq$10~hr.  Two sources were identified as ideal initial targets (Table~\ref{tab:bdProp}).

2MASS~J08354256$-$0819237 (hereafter {\sha}; \citealt{2003AJ....126.2421C}) is a bright, nearby ($d$ = 7.27$\pm$0.02~pc; \citealt{2016AJ....152...24W}) L5 dwarf that shows signatures of low surface gravity \citep[][however, see \citet{2015ApJS..219...33G}]{2016ApJ...833...96L}. 
There is no Li I absorption reported for this source \citep{2003AJ....126.2421C,2015ApJS..220...18B}; however its low temperature \citep[T$_{eff}$=1800~K][]{2015ApJS..219...33G} and likely youth ($<$1 Gyr) is consistent with a mass below the hydrogen-burning mass limit \citep{2003A&A...402..701B,2000ApJ...531..438B}.
This source has been reported to vary in both optical and near-infrared bands with a period of 3.1~hr and an amplitude of $\sim$1\%  \citep{2004MNRAS.354..378K,2014A&A...566A.111W}.
Its $v\sin{i}$ = 14.2$\pm$0.4~{\kms} \citep{2010ApJ...723..684B} suggests a near pole-on viewing angle assuming a radius\footnote{We adopt the equatorial radius of Jupiter, {\rjup} = 7.15$\times$10$^9$~cm.} R = 1~{\rjup}.
There is an $H$ = 12.3 reference star (2MASS~J08354383$-$0819212) at a separation of 28$\arcsec$ suitable for relative flux calibration (Figure~\ref{fig:images}). 

2MASS J18212815+1414010 (hereafter {\shb}; \citealt{2008ApJ...686..528L}) is a similarly bright and nearby ($d$ = 9.38$\pm$0.03; \citealt{2016MNRAS.455..357S}) L4.5 dwarf, which exhibits a peculiar near-infrared spectrum attributed to youth and/or thick clouds \citep{2008ApJ...686..528L,2015ApJS..219...33G,2016ApJ...833...96L} and Li absorption \citep{2008ApJ...686..528L}.
This source is a photometric variable with a period of 4.2$\pm$0.1~hr \citep{2015ApJ...799..154M} and $v\sin{i}$ = 28.85$\pm$0.16~{\kms}  \citep{2010ApJ...723..684B}, suggesting a near equator-on orientation.  \citet{2015ApJ...798L..13Y} measured 1--3\% spectral variability with $HST$/WFC3 over 1.1--1.6~$\micron$, with greater variabilty at shorter wavelengths.
There are three suitably bright reference stars within 40$\arcsec$ for relative calibration visible in Figure \ref{fig:images}.
We chose 2MASS~J18212622+1414064 (visible in Figure~\ref{fig:images}) with $H$=12.0 at 32$\arcsec$ separation for our monitoring observations because this configuration minimized the amount of light entering the slit from the nearby visual companion to the north of our target.
However, we expect that another configuration with the reference star to the southwest could achieve similar spectro-photometric precision.

\begin{figure*}
\centering
	\plottwo{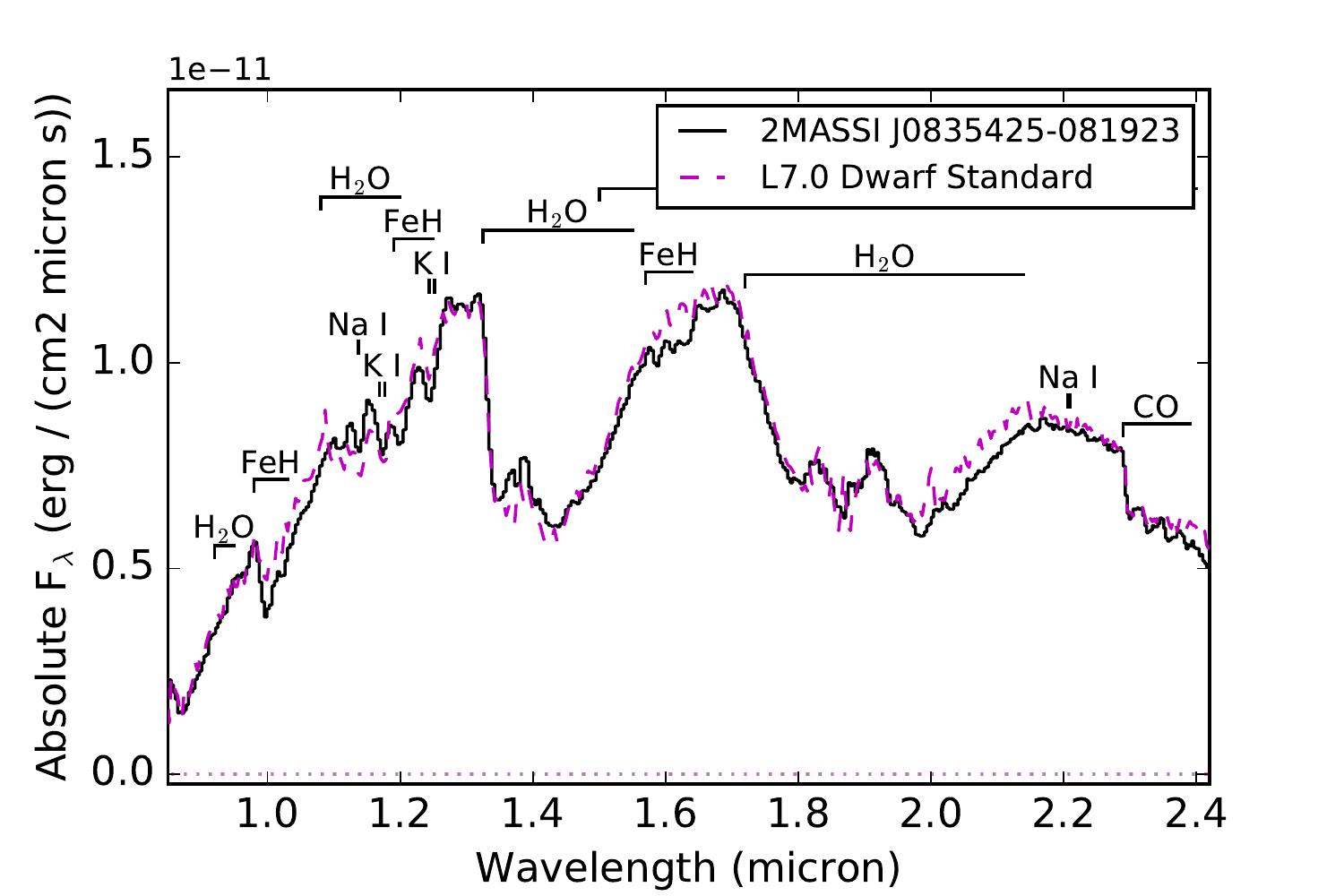}{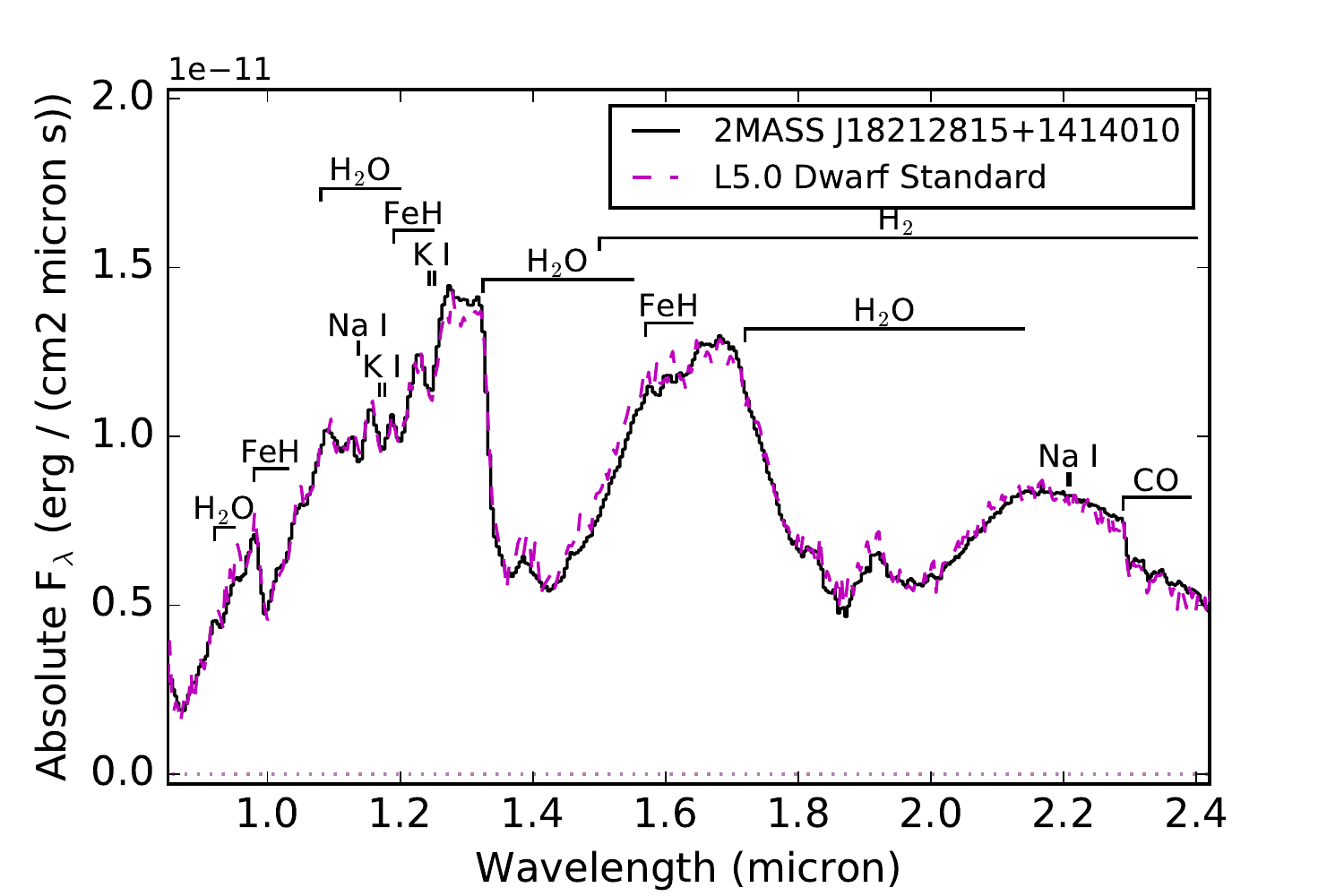}
	\plottwo{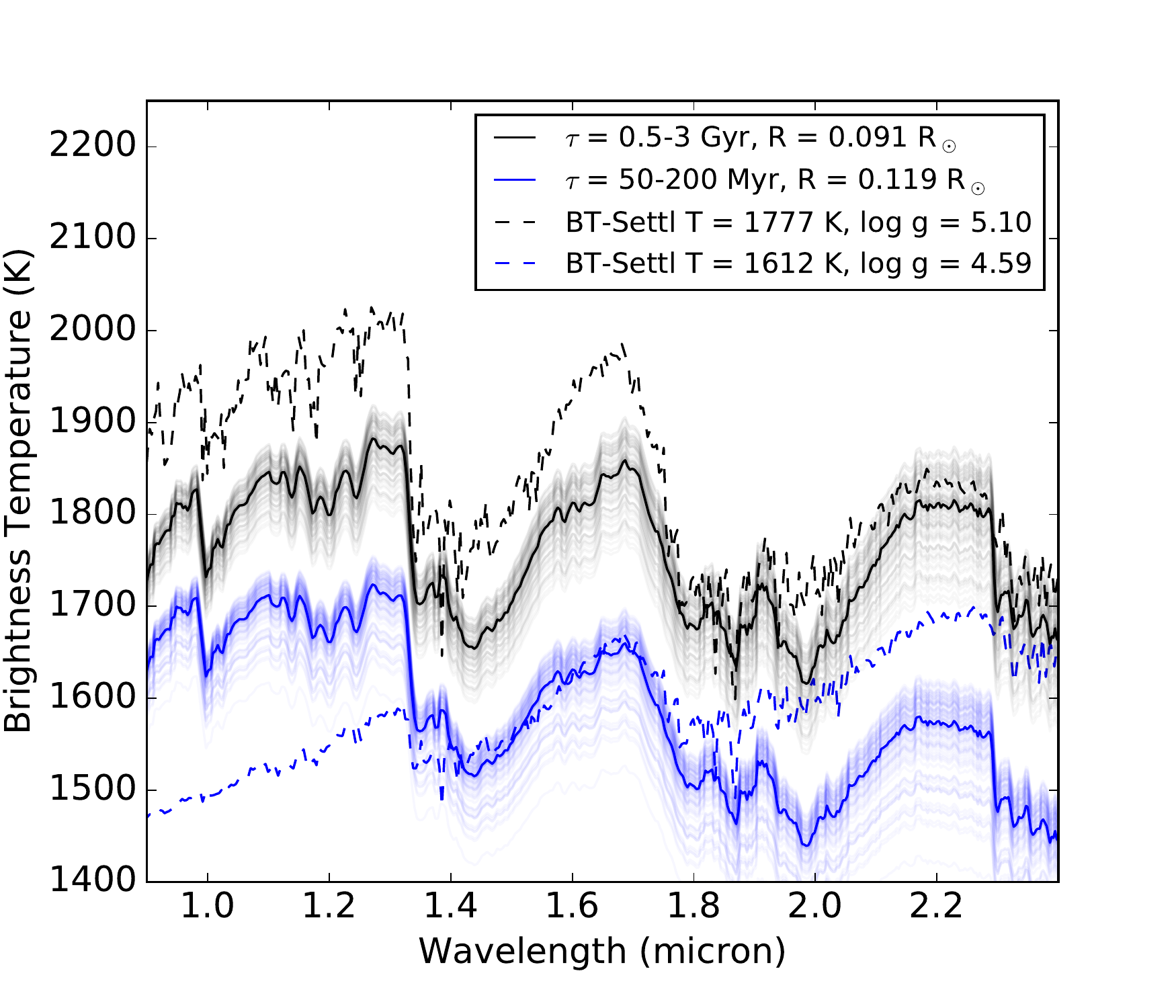}{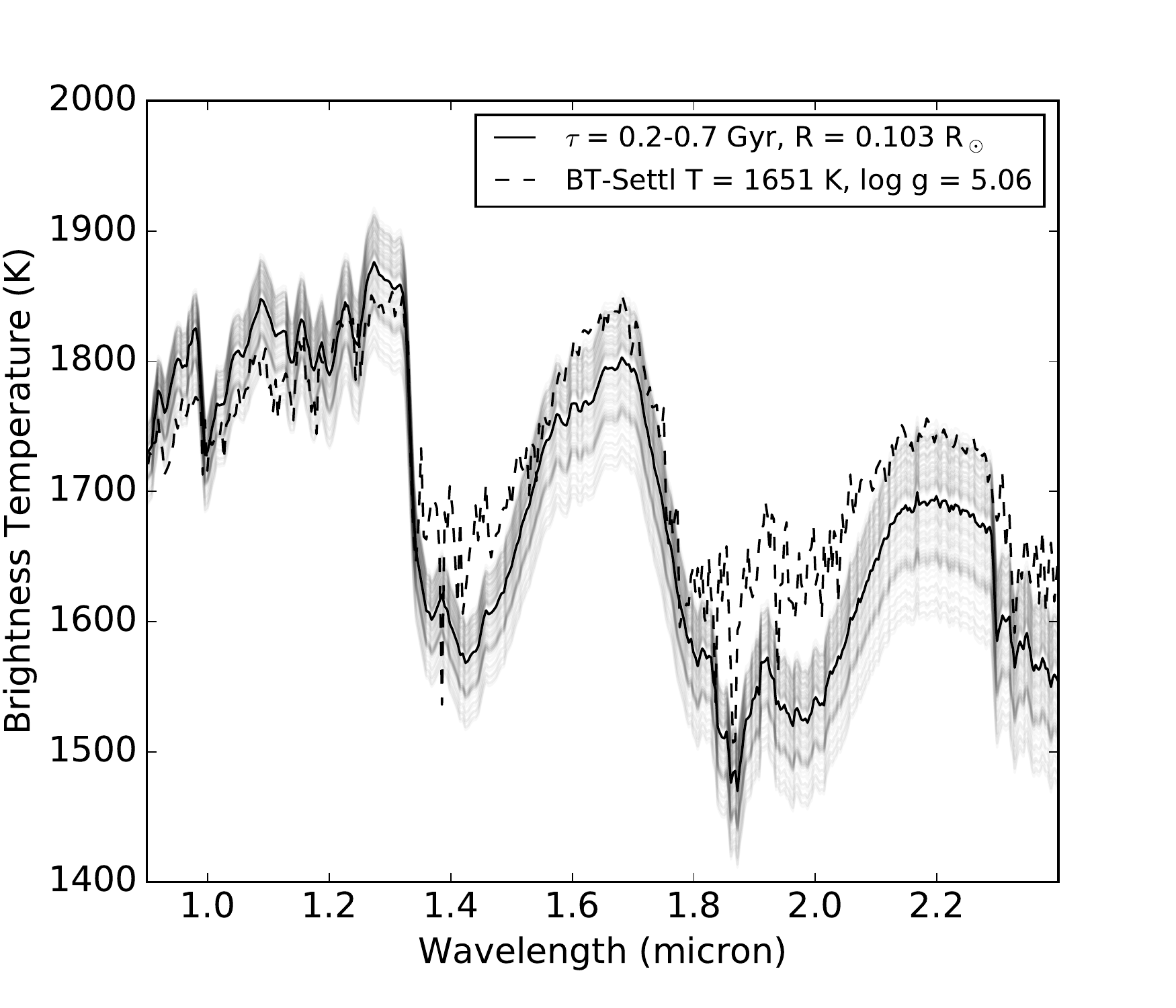}
	\caption{(Top) SpeX prism spectra (in black) of {\sha} (left; \citealt{2010ApJ...710.1142B}) and {\shb} (right; \citealt{2008ApJ...686..528L}) scaled to absolute flux densities, compared to their closest-match L dwarf spectral standards as defined in \citet[in magenta]{2010ApJS..190..100K}: the L7 2MASSI~J01033203+1935361 (data from \citealt{2014ApJ...794..143B}) and the L5 SDSS~J083506.16+195304.4 (data from \citealt{2006AJ....131.2722C}), respectively. 
	Primary absorption features are labeled.
	(Bottom) Calculated brightness temperature spectra of {\sha} (left) and {\shb} (right).
	The faint lines represent individual calculations for a Monte Carlo realization of spectral flux, absolute magnitude, {\lbol} and age uncertainties; the dark lines represent the median of these realizations.
	For {\sha}, we show two  brightness temperature spectra corresponding to a young brown dwarf (50--200 Myr; blue) and a field dwarf (0.5--3~Gyr; black).
	We also show the brightness temperature spectra of the equivalent spectral model for the measured luminosity and assumed ages (dotted lines).
	The corresponding {\teff} and {\logg} values are given in the legend.}
	\label{fig:tbright}
	\vspace{0.1in}
\end{figure*}

To characterize the atmospheric regions probed by our observations, we analyzed previously 
obtained IRTF/SpeX data for these sources \citep{2008ApJ...686..528L,2010ApJ...710.1142B}.
Direct comparison to spectral standards following the convention of \citet{2010ApJS..190..100K} indicates near-infrared
classifications of L7 for {\sha} and L5 for {\shb} (Figure~\ref{fig:tbright}). The former is considerably later than the L5 optical classification previously reported by \citet{2003AJ....126.2421C}. We confirm the intermediate low surface gravity classification (INT-G) reported by \citet{2016ApJ...833...96L} based on the indices of \citet{2013ApJ...772...79A} using an L7 classification. The discrepancy between the optical and near-infrared classifications of this source is likely related to surface gravity effects. {\shb}, on the other hand, is a good match to the L5 spectral standard and has a field dwarf gravity classification.

We computed brightness temperature spectra from these data as
\begin{equation}
T_{\lambda,surf} = \frac{hc}{k\lambda}\left[\ln\left(1+\frac{2\pi{hc^2}}{F_{\lambda,surf}\lambda^5}\right)\right]^{-1},
\end{equation}
where the surface flux density $F_{\lambda,surf}$ is computed as
\begin{equation}
F_{\lambda,surf} = {\pi}B_{\lambda} = F_{\lambda,obs}\left(\frac{d}{R}\right)^2 = F_{\lambda,abs}\left(\frac{10~{\rm pc}}{R}\right)^2.
\end{equation}
$F_{\lambda,obs}$ and $F_{\lambda,abs}$ are the observed and absolute flux densities, $B_\lambda$ is the Planck function, $d$ is the source distance and $R$ is the source radius.
We have ignored limb darkening in our calculations, which affects the surface intensity of brown dwarf surfaces \citep[e.g.][]{2011A&A...529A..75C}.
Since there are few empirical constraints on limb darkening for L-dwarfs, we assume a uniform brightness temperature surface, which gives an average brightness temperature of a non-homogeneous surface.

Both spectra were calibrated to absolute flux densities using their 2MASS $H$ magnitudes and measured parallaxes. We adopted radii from the evolutionary models of \citet{2003A&A...402..701B} assuming bolometric luminosities from \citet{2015ApJ...810..158F} for {\sha} and from\footnote{The parameters reported for {\shb} in \citet{2016ApJS..225...10F} are in error, but have been updated here based on a forthcoming erratum (J.\ Faherty, 2017, priv.\ comm.).} \citet{2016ApJS..225...10F} for {\shb} (Table~\ref{tab:bdProp}). Uncertainties in the spectral fluxes, absolute photometric calibration, adopted {\lbol} and radii were propagated by Monte Carlo simulation.  For {\sha} we considered two cases: radii of 0.114--0.137~R$_{\odot}$ for ages of 50--200~Myr, the age range estimated for INT-G sources by \citet{2013ApJ...772...79A}; and radii of 0.088--0.101~R$_{\odot}$ for ages of 0.5--3~Gyr; i.e., a field dwarf age.
For {\shb}, we assumed a lower age limit of 200~Myr based on its field gravity classification, and an upper limit of 700~Myr based on the analysis of \citet{2016MNRAS.455..357S}. The \citet{2003A&A...402..701B} evolutionary models predict radii of 0.097--0.114~R$_{\odot}$ for these ages and the luminosity of {\shb}.  

Figure~\ref{fig:tbright} shows the resulting brightness temperature spectra and their uncertainties from these calculations.  The spectral peaks for {\sha} are around 1800--1850~K in the field surface gravity case, and decline from 1700~K at $J$-band (1.2~$\micron$) to 1580~K at $K$-band (2.2~$\micron$) in the low surface gravity case.
We compared these two cases to solar-metallicity BT-Settl spectral models from \citet{2012RSPTA.370.2765A}, with {\teff} and {\logg} parameters consistent with the bolometric luminosity of {\sha} and the adopted ages (again based on the \citealt{2003A&A...402..701B} evolutionary models).
To achieve temperatures and gravities between the pre-computed spectra in the BT-Settl atmospheric grid, we interpolated the spectra.
Both models overlap the data to some degree, although the young model shows an increasing brightness temperature with wavelength rather than a decreasing temperature.  These comparisons slightly favor the field surface gravity case, although differences in the properties of dust between source and models likely play a significant role. For {\shb}, 
brightness temperatures range from 1850~K at $J$-band to 1670~K at $K$-band, again overlapping but declining more strongly with wavelength than the equivalent spectral model, which has a {\teff} similar to that inferred by \citet{2016ApJS..225...10F} and the  \citet{2015ApJ...810..158F} {\teff}/spectral type relation.

\section{Observations}\label{sec:obs}
\begin{deluxetable*}{lcccccccc}[b]
\tablecaption{Observations\label{tab:obsParam}}
\tablewidth{0pt}
\tablehead{
 & \multicolumn{2}{c}{Observation Epoch (UT)} \\
 \cline{2-3}
\colhead{Target} & 
\colhead{Date} & 
\colhead{Start} & 
\colhead{$N_{int}$} & 
\colhead{$t_{int}$} & 
\colhead{Duration} & 
\colhead{Airmass} & 
\colhead{Reference Star} & 
\colhead{FWHM \@ $H$} \\
 & &  & & \colhead{(s)} & \colhead{(hr)} & \colhead{Start-End} & & \colhead{(\arcsec)}}
\startdata
{\sha} & 2015-12-31 & 09:35:10 & 272 & 45 & 4.65 & 1.59-1.28 & J08354383-0819212 & 0.5 \\
{\shb} & 2016-06-25 & 07:29:31 & 352 & 60 & 7.47 & 1.42-2.15 & J18212622+1414064 & 0.5 \\
\enddata
\tablecomments{N$_{int}$ is the number of integrations and t$_{int}$ is the effective integration time for the 32 non-destructive reads up the ramp. The duration is the total time from start to end of the observations including array sampling and overheads. The FWHM is the measured full width at half maximum of the spatial point spread function around H band (1.7$\mu$m), taken as the median value over the entire time series.}
\end{deluxetable*}

We observed {\sha} and {\shb} for one night each with the IRTF/SpeX (Table \ref{tab:obsParam}).
{\sha} was observed for 4.65 hours, covering 1.5 rotation periods; {\shb} was observed for 7.47 hours, covering 1.8 rotation periods.
Following \citet{2014ApJ...783....5S}, we used SpeX in its prism-dispersed mode, obtaining 0.9--2.4~$\micron$ spectra in a single order at low dispersion. 
Both target and reference star were observed simultaneously by orienting the 3$\arcsec$$\times$60$\arcsec$ slit to align with the visual binary axis. This orientation is generally misaligned with the parallactic angle, but the good seeing (0$\farcs$5 at $H$-band) and wide slit minimizes differential color refraction. The effective resolution of the spectral data, set by seeing and pointing errors, was $\lambda$/$\Delta\lambda$ $\approx$ 50. Integration times were 45~s and 60~s for {\sha} and {\shb}, respectively. The short-wavelength limit of our spectral coverage was set by the use of a 0.9~$\mu$m dichroic which deflected optical light into the MIT Optical Rapid Imaging System (MORIS) camera \citep{2011PASP..123..461G}, which was used for $i$-band guiding.


\begin{figure*}[!t]
\centering
\subfloat[2MASS J0835]{
	\includegraphics[width=0.5\textwidth]{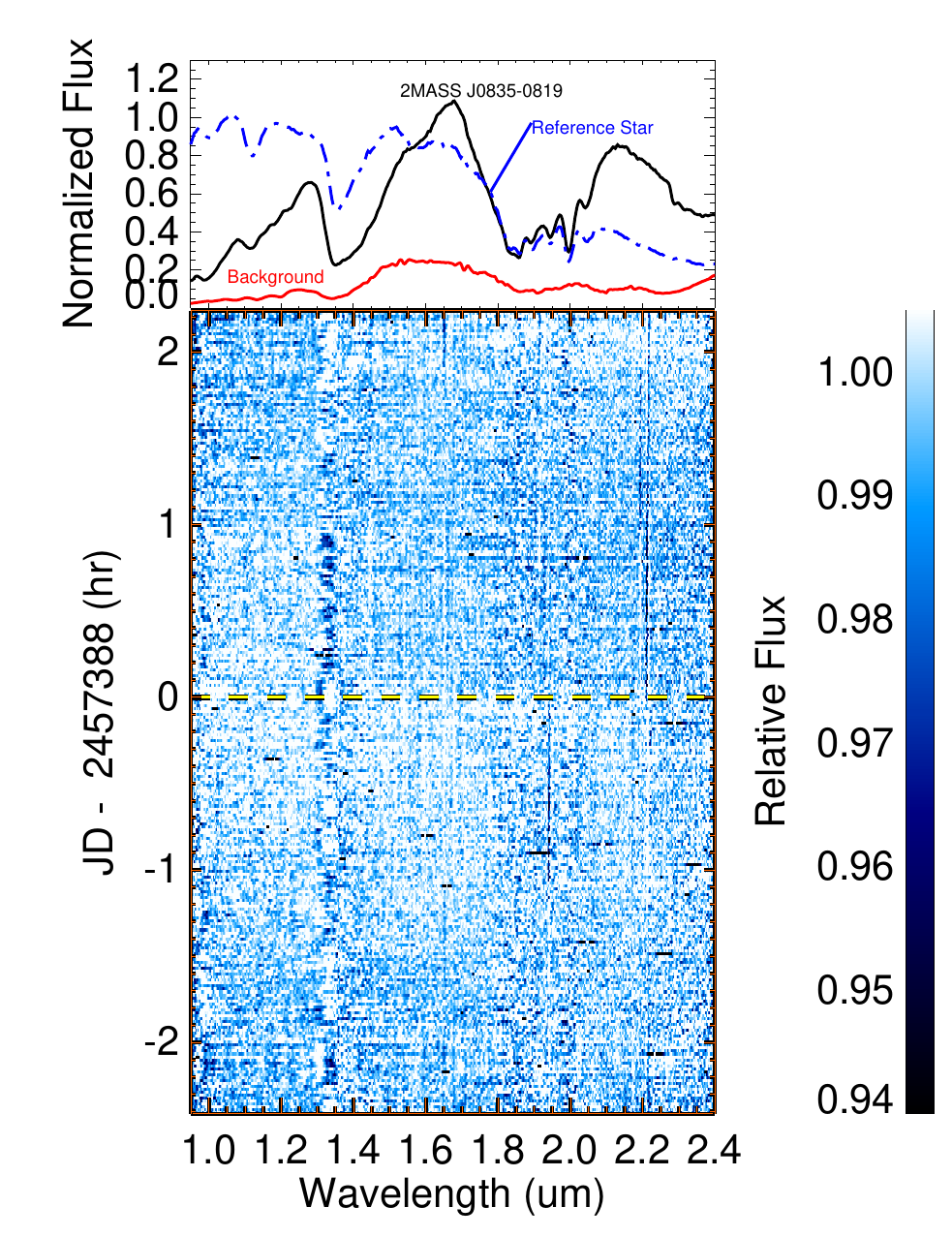}
	\label{fig:specphot0835}
	}
\subfloat[2MASS J1821]{
	\includegraphics[width=0.5\textwidth]{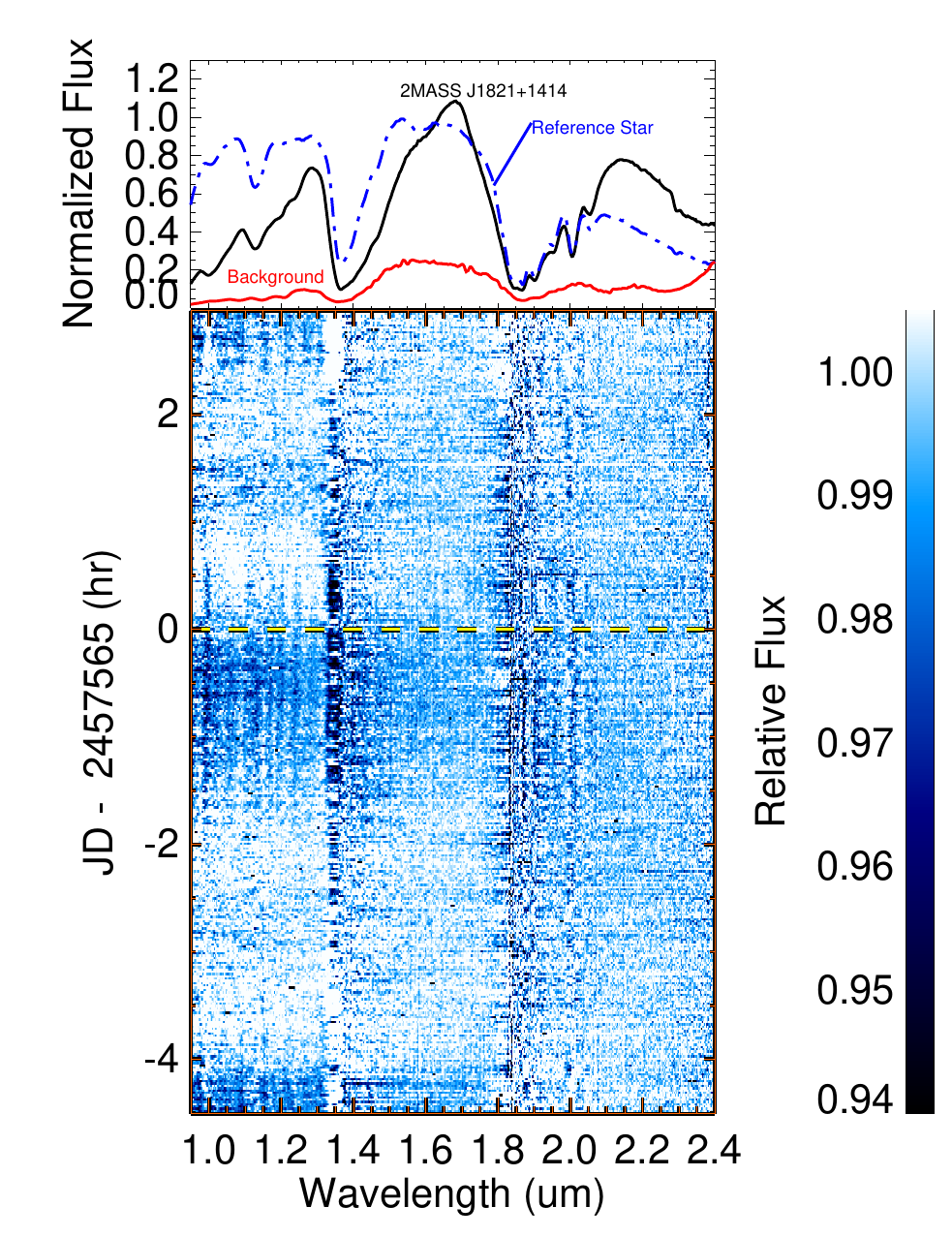}
	\label{fig:specphot1821}
	}
	\caption{Top Panels: Normalized median raw spectra of {\sha} (left) and {\shb} (right) in black, compared with the respective reference stars (blue) and background emission (red; mostly telluric OH emission and thermal background). The areas of higher telluric absorption are visible near 1.4~$\mu$m and 1.8~$\mu$m.  Bottom Panels: Dynamic spectra of each brown dwarf system. The dynamic spectrum for \sha\ shows no significant variability in the spatial or spectral direction whereas \shb\ shows significant modulations with a period near 4 hours.
The variability in \shb\ is prominent at the short wavelengths with a magnitude of $\sim$2--3\% peak-to-trough, but $\lesssim 0.5\%$ toward 2~$\mu$m.}
	\label{fig:specphot}
	\vspace{0.1in}
\end{figure*} 

The 2D prism images, which have a dispersion direction approximately along the X direction and spatial direction approximately along the Y direction, were rectified so that the X coordinates of all background spectra corresponded to the same wavelength.
This was achieved by selecting a reference row between the sources and then cross-correlating this reference spectrum with all other illuminated detector rows.
A third order polynomial was fit to the cross correlation peaks (masking out the sources) to determine a one dimensional smooth X (spectral) shift solution as a function of Y (spatial) position  \citep{2016ApJ...826..156S}.
Once the 2D spectral images were rectified, the background is subtracted along the Y (spatial) direction with a 4th order polynomial.
The reduction software is available online\footnote{https://github.com/eas342/spectral\_pipeline\_and\_scripts}.
 
We obtained ThAr arc lamp spectra, dark frames and flat field data for calibration.
The ThAr data is taken for wavelength calibration with the 0.3$\arcsec\times60\arcsec$ slit since it has a resolving power high enough to separate the ThAr lines and identify them.
This wavelength calibration data is taken at the very beginning and very end of the sequence to ensure the 3$\arcsec \times 60 \arcsec$ slit is unmoved during the rest of the calibration and science data.
Dark frames are taken with the same exposure time as the science images and with the order sorting filter set to the closed position.
We took flat field field data with an incandescent lamp to correct for high frequency, pixel-to-pixel variations.
For {\sha}, we used a combination of the flat-field lamp and a sky frame to construct a pixel response map, removing large-scale ($\gtrsim$ 30 pixels) structure with low order polynomial fits along spatial and spectral axes, and normalizing to the median response spectrum. 
To correct for instrument flexure, we cross-correlated the sky frame with each science image to measure linear spatial shifts, which were $\pm$5~pixels or 0.5\arcsec\ for \sha\ and $\pm$1.5~pixels or 0.15\arcsec\ for \shb.
These shifts were used to apply a custom flat field for every image: the illumination flat (with pixel-to-pixel flat removed) is shifted to the cross correlation peaks and multiplied by the pixel-to-pixel flat structure, which is maintained at the same position.
For {\shb}, we did not have sufficient sky frames so we used our incandescent lamp image for both flat fielding and flexure analysis.

As described in \citet{2016ApJ...826..156S}, we performed spatial background subtraction over the entire slit using a 4th-order polynomial fit to points more than 30 pixels away from the center of the source dispersion traces.  Spectra were optimally extracted \citep{1986ApJ...302..757H}, with spatial profiles fit to low order polynomials along the spectral direction one row at a time with 20$\sigma$ rejection to remove cosmic rays and bad pixels.
These spatial profiles were then used to calculate a weighted sum spectrum across the illuminated detector for each source and exposure. 
Typical signal-to-noise of the extracted spectra were 40 - 100 per spectral pixel (unbinned).
Relative spectra (source divided by reference) were then computed for each exposure.
Our image rectification process described above ensures that the wavelength solution is the same for each row of the image.
However, there are small additional shifts of the sources within the slit, so another shift was performed between the sources to ensure that telluric features were removed when dividing the two spectra.


\begin{figure*}[!t]
\centering
\subfloat[2MASS J0835]{
	\includegraphics[width=0.5\textwidth]{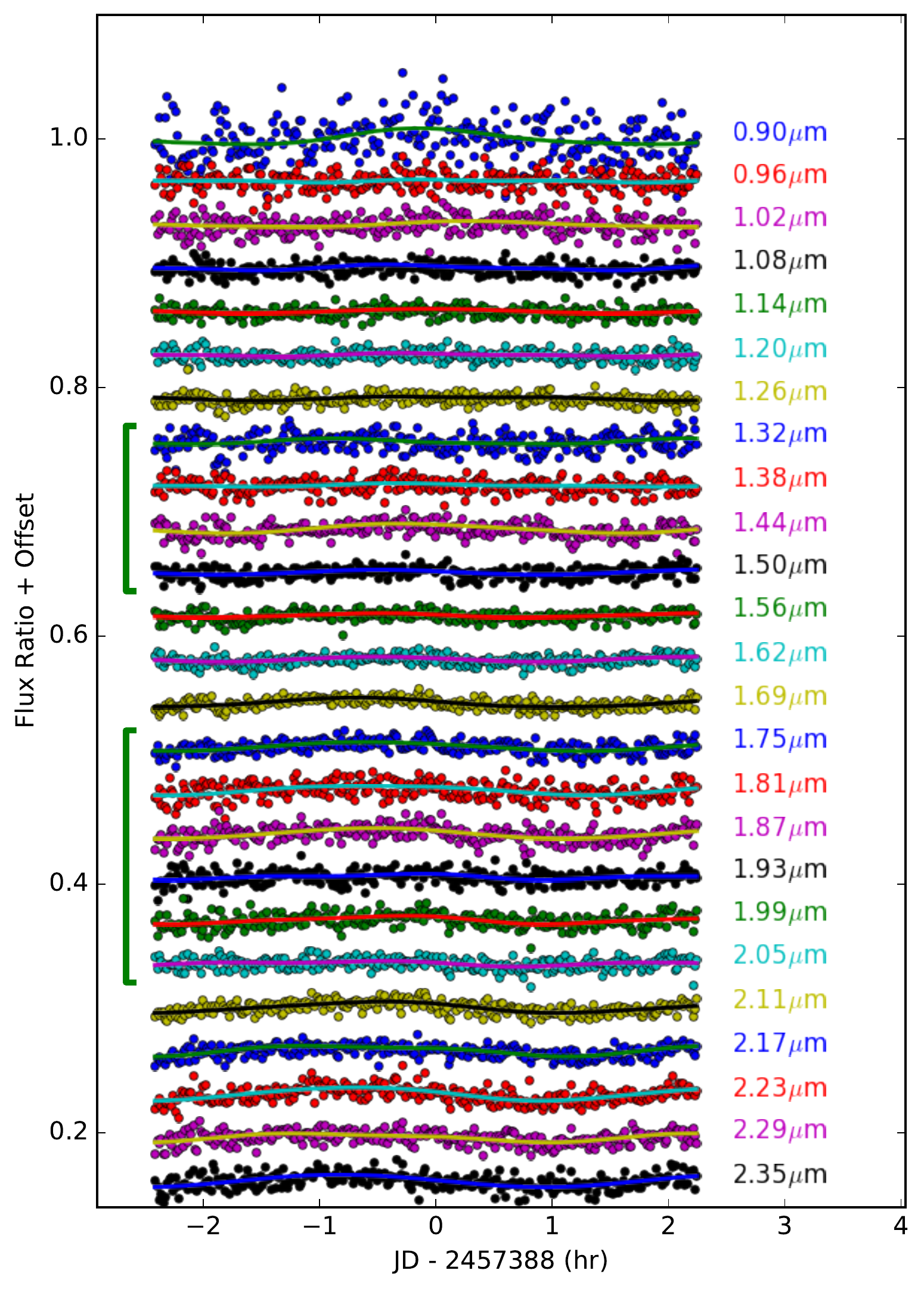}
	\label{fig:tserPhase0835}
	}
\subfloat[2MASS J1821]{
	\includegraphics[width=0.5\textwidth]{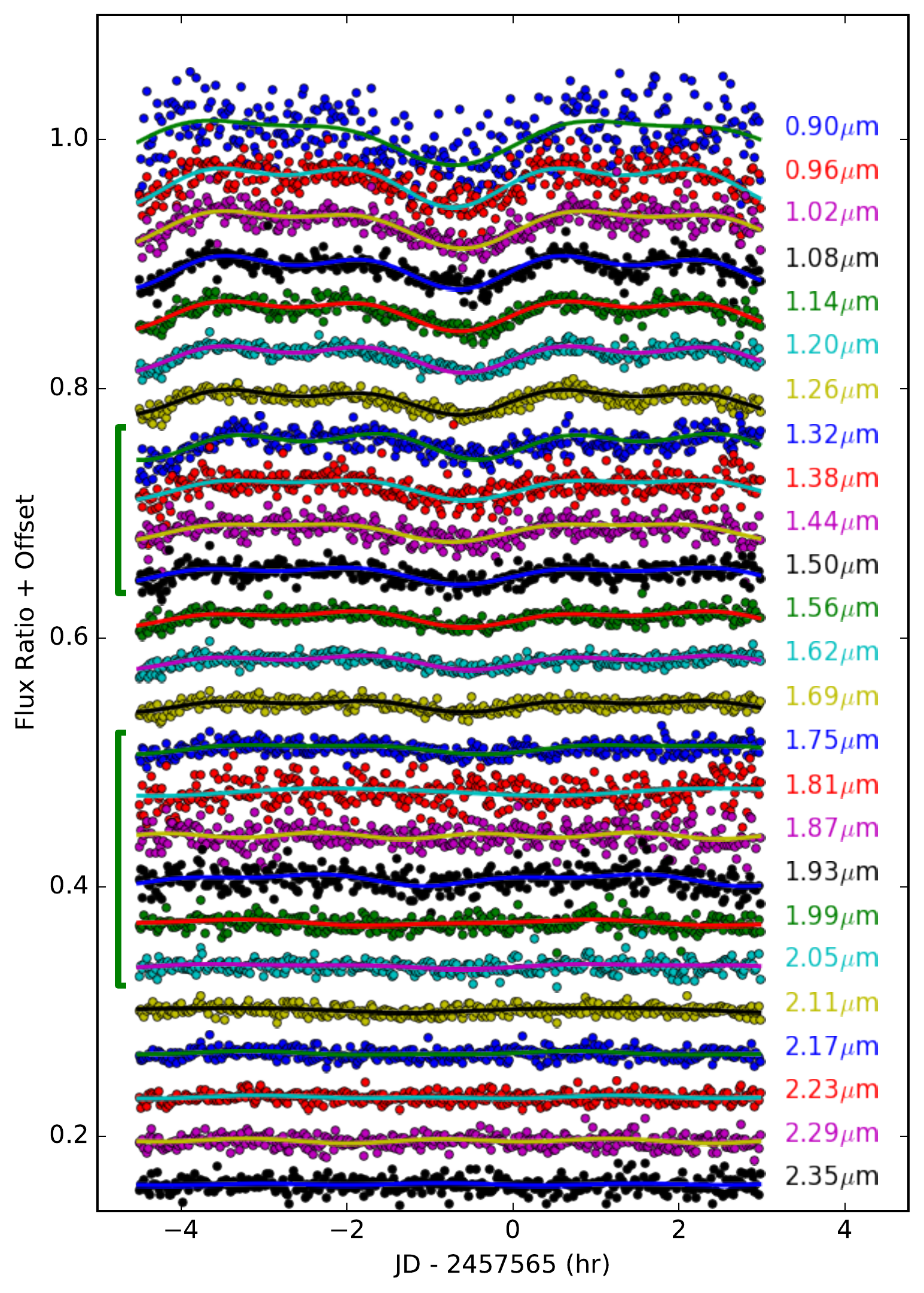}
	\label{fig:tserPhase1821}
	}
	\caption{Time series for each system, after subtraction of baseline trends. Each wavelength binned lightcurve is color coded, offset, and labeled on the right side. \sha\ shows relatively flat light curves whereas \shb\ has clearly visible double-peaked oscillations at short wavelengths and flat light curves at long wavelengths.
	Lines trace the best double sinusoidal model fit for each light curve.
	Green brackets show regions where telluric absorption may contaminate the time series.}
	\label{fig:tserDetrend}
\end{figure*}

\begin{figure*}
\centering
\subfloat[2MASS J0835]{
	\includegraphics[width=0.5\textwidth]{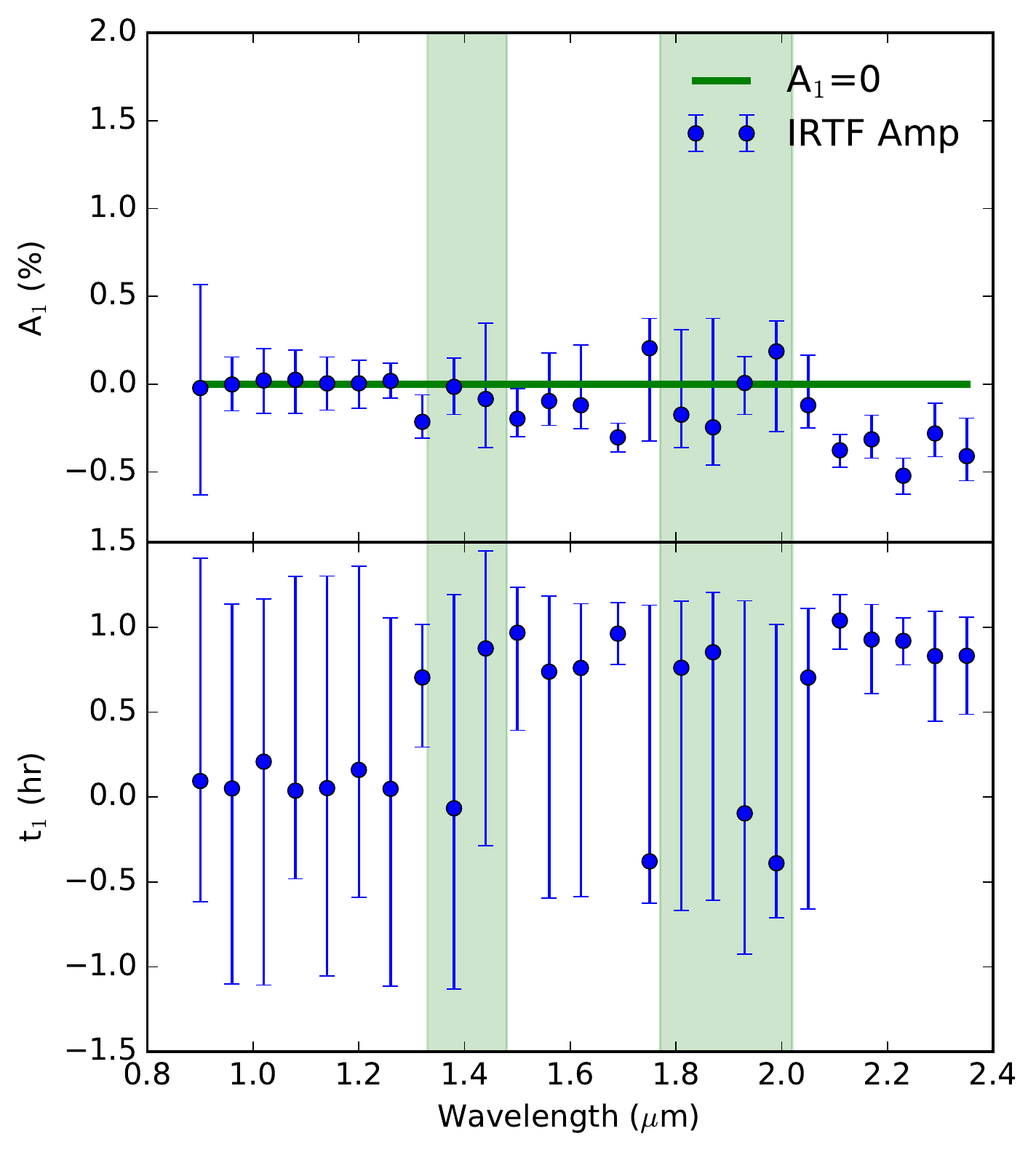}
	\label{fig:ampspec0835}
	}
\subfloat[2MASS J1821]{
	\includegraphics[width=0.5\textwidth]{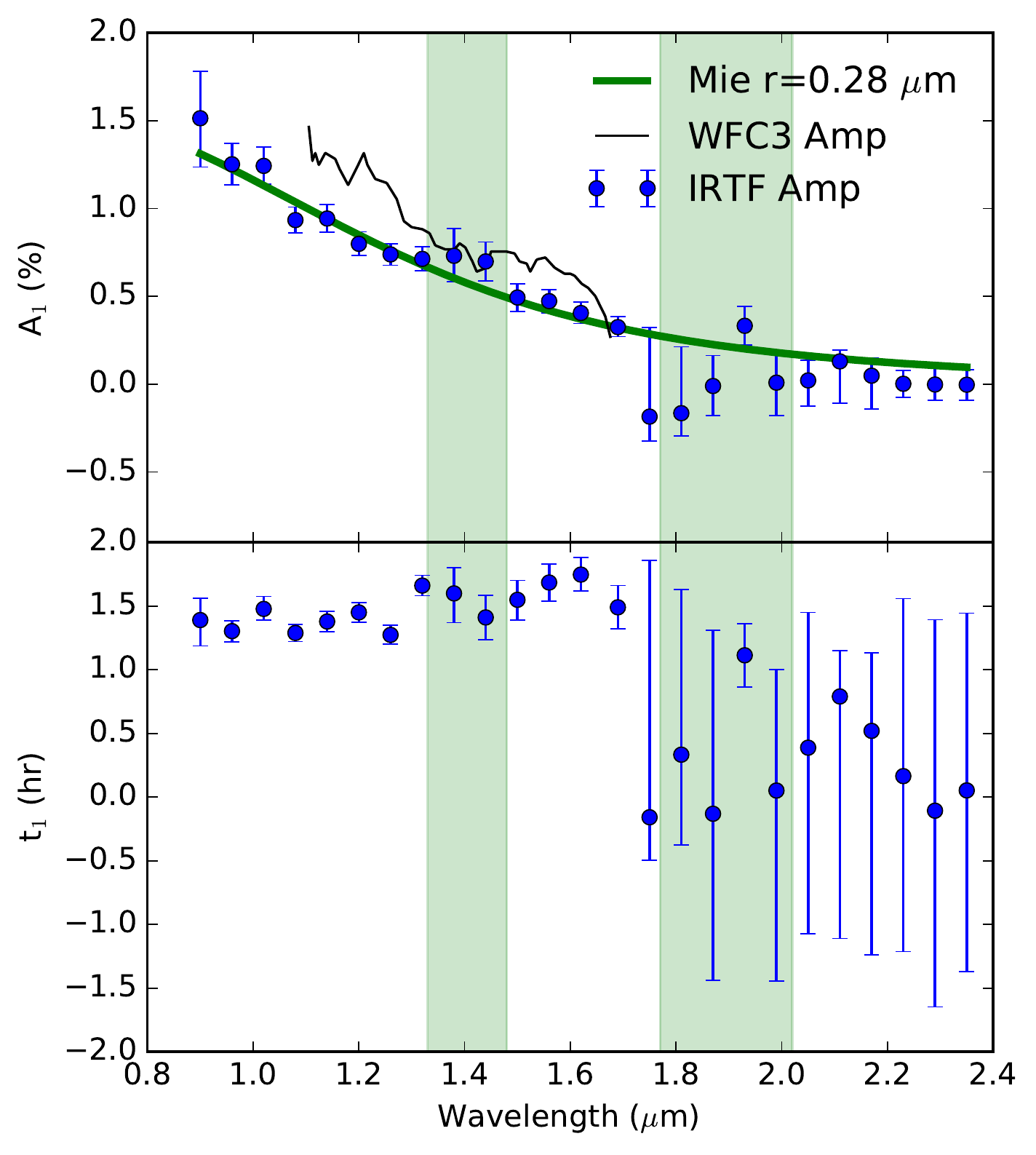}
	\label{fig:ampspec1821}
	}
	\caption{(Top panels) Sine fit amplitude versus wavelength for each brown dwarf. \sha\ shows only marginally significant variability near 1.7~$\micron$ and 2.1--2.4~$\micron$, and no obvious trends with wavelength. \shb\ has significant variability over wavelengths $<$1.7$\micron$ with a strong spectral slope.
	 Also shown is the WFC3 amplitude spectrum of \shb\ from \citet{2015ApJ...798L..13Y}. The WFC3 observations were obtained at a different epoch, but show a similar slope.
	 The amplitude spectrum for \shb\ is fit to a simple Mie scattering model with a log-normal particle size distribution of spherical forsterite grains; the best fit has a median particle radius of 0.28 $\mu$m.
	 In all plots the regions potentially contaminated by telluric absorption are shaded in light green.
	 {\it Bottom Panels:} The time offset of maximum amplitude, $t_1$ (in hours), relative to the reference epoch (JD 2457388.0 for \sha\ and JD 2457565.0 for {\shb}).
	 Neither shows a clear trend with wavelength, indicating no significant phase offsets.
	 For wavelengths with small variability amplitudes, $t_1$ becomes as wide as the prior window $-\tau/2 < t_1 < \tau/2$.}
	\label{fig:ampSpec}
\end{figure*} 

\begin{figure}
\begin{centering}
\includegraphics[width=0.5\textwidth]{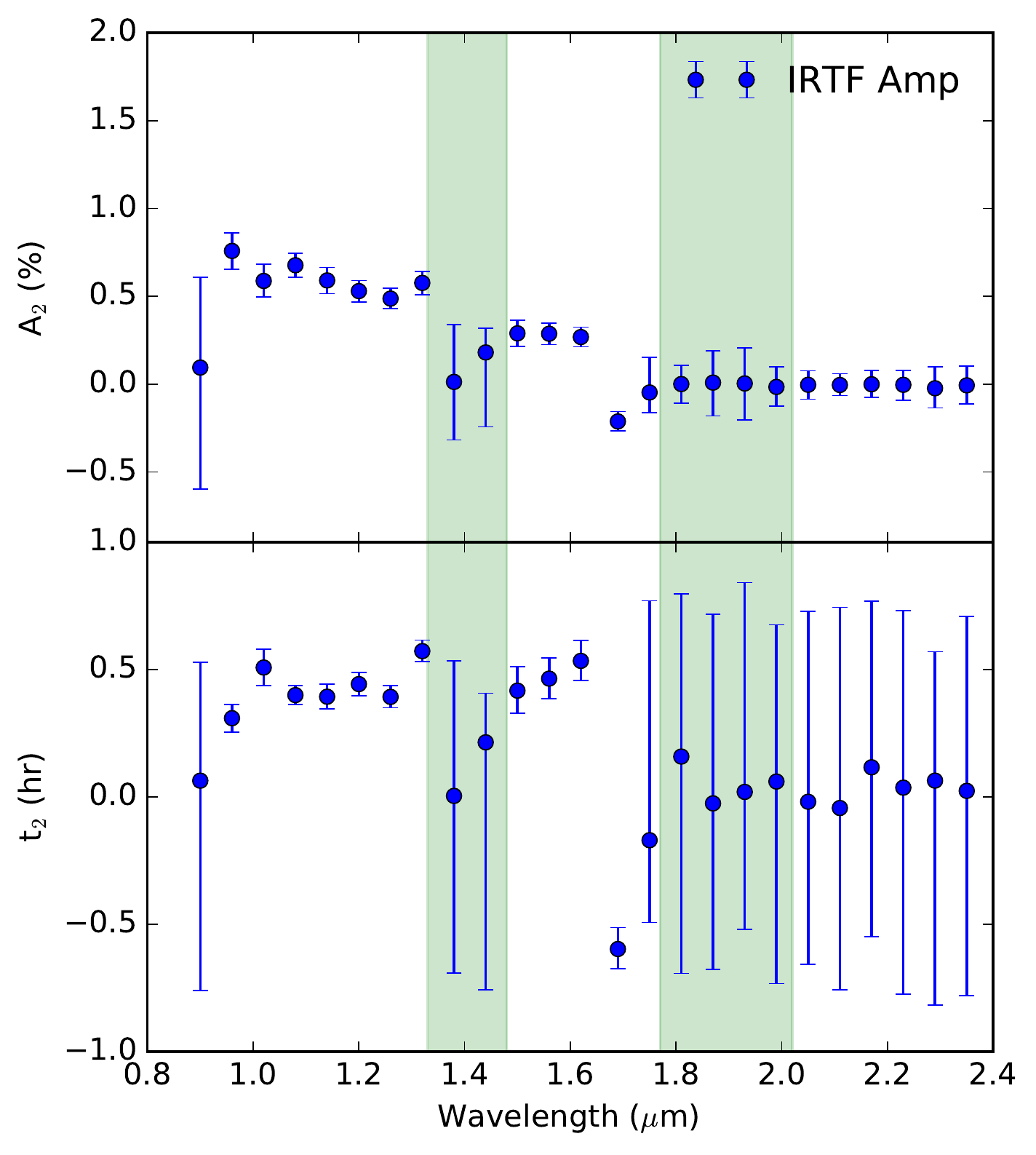}
\caption{Secondary sinusoidal amplitude ($A_2$) as a function of wavelength for \shb.
As in Figure \ref{fig:ampSpec}, the regions potentially contaminated by telluric absorption are shown in green.
The secondary amplitude $A_2$ shows a gradual decrease as a function of wavelength, similar to that seen with $A_1$.
{\it Bottom Panel:} The time offset of maximum variation, $t_2$ (in hours) shows no clear phase shift trend.
The constraints approach the prior $\pm \tau$/4 when no significant variability is detected.
The $A_2$ constraints for \sha\ (not shown) are all consistent with zero.}\label{fig:amp2Spec}
\end{centering}
\end{figure}

\section{Analysis}\label{sec:analysis}

\subsection{Light Curves}

Figure \ref{fig:specphot} shows the raw spectra of the two brown dwarfs, their reference spectra and the background spectra.
Neither the reference spectra nor the brown dwarf spectra have had any corrections applied for telluric absorption or instrumental spectral response.
The reference stars have significantly bluer spectra than the brown dwarfs, but nonetheless contain the same telluric water absorption signatures
near 1.37~$\mu$m and 1.83~$\mu$m, resulting in significantly lower flux.

Figure \ref{fig:specphot} also shows the dynamic spectra, the relative source brightness as a function of wavelength and time.
{\sha} has dynamic spectra resembling white noise, with no significant variations in either spectral or temporal axes at the $\lesssim$0.5\% level.
{\shb}, on the other hand, shows a clear periodic signal that is strongest at the shortest wavelengths and declines in amplitude toward the longest wavelengths. The periodic signal is nearly coincident with the 4.2~hr period measured by \citet{2015ApJ...799..154M}, enabling confirmation that the signal is astrophysical rather than systematic (see \citealt{2016ApJ...826..156S}).
Both sources exhibit some systematic features in the strong telluric absorption bands at 1.4 and 1.8~$\mu$m, likely caused by residual wavelength drifts from flexure and lower signals-to-noise.

Individual narrow-band light curves are extracted for each target in 25 spectral bins of width 0.060~$\mu$m; these are shown in Figure~\ref{fig:tserDetrend}.
The baseline trends (which scale linearly with time and linearly with airmass) have been removed from Figure~\ref{fig:tserDetrend}, but the original light curves can be viewed in Figure \ref{fig:tserNoDetrend}.
The uncertainties in these curves due to photon and read noise range from 0.2\% to 0.6\% per exposure.
We measure the standard deviation of the time series for wavelength regions that have no significant variations near the rotation period for each brown dwarf and compare it to the expected photon and read noise.
In these wavelength bins, the standard deviations of the points are about 4$\times$ the expected read noise and photon noise.
We attribute this to correlated errors between wavelengths and as a simple approximation multiply all error bars by a factor of 4 to account for these systematics.

We find minimal variation in the lightcurves of {\sha} but pronounced variation in {\shb}, particularly in bands $<$1.8~$\mu$m. 
These variations are periodic with ``double-hump'' structures at peak amplitudes around -2.7~hr and +1.3~hr relative to JD 2457565.00.
The amplitude of these variations is strongly wavelength-dependent with $\sim$3.0\% peak-to-trough variations at short 
wavelengths and near-zero amplitude at the longest wavelengths.

Each of the raw light curves was fit to a double sinusoidal model:
\begin{equation}\label{eq:cosfit}
\begin{split}
F(t) = \left( A_1 \cos(2 \pi (t - t_1)/\tau) + A_2 \cos(4 \pi (t - t_2)/\tau) + 1\right) \\
\times (B t + C + D a(t))
\end{split}
\end{equation}
where $F(t)$ is the relative flux as a function of time $t$, $A_1$ and $A_2$ are the amplitudes of the two sinusoid terms, $t_1$ and $t_2$ are the time of maximum periodic flux for each sinusoid term, $\tau$ is the period of the variations, $a(t)$ is the airmass as a function of time and $B$, $C$ and $D$ are constants quantifying baseline (i.e., non-astrophysical) trends.
Since some wavelengths are affected by telluric contamination and we wish to avoid fitting harmonics, we impose a prior on the rotational period $\tau$.
We impose 2.6 $< \tau <$ 3.6 hr for \sha\ and 3.7 $< \tau <$ 4.7 hr for {\shb}, $\pm$0.5 hr around the literature values.
These priors are wide enough they don't drive the constraints on $\tau$ at short wavelengths for {\shb}, but narrow enough to ensure that the sinusoidal fits do not just fit the airmass trends with time.
We also restrict the time of maximum periodic flux such that $-\tau/2 < t_1 < \tau/2$ and $-\tau/4 < t_2 < \tau/4$ to assure there are no 2$\pi$ degeneracies in the fit.
We do not put any constraints on A$_1$ relative to A$_2$.
The co-variance between A$_1$ and A$_2$ is small ($\lesssim 10^{-2}$ of variance of A$_2$ in absolute magnitude) resulting in ``round'' MCMC 2D posterior distributions between the two parameters since they are essentially different Fourier parameters for the time series.

Figure \ref{fig:ampSpec} shows the MCMC posterior constraints for periodic amplitudes as a function of wavelength.
For {\sha} (Figure \ref{fig:ampspec0835}) we find at most small amplitude variability ($\lesssim 0.5\%$), marginally significant at 1.7~$\mu$m and 2.1--2.35~$\mu$m, with a flat trend as a function of wavelength.
If one were to average all of the K-band points together, they are consistent with a 3.4 $\pm$ 0.1 hr rotation period, though this is somewhat driven by the prior 2.6 $< \tau < $3.6 hr.

\shb, on the other hand, shows highly significant variability with a decreasing trend in amplitude as function of wavelength.
In both cases, we find no statistically significant wavelength dependence on the time of maximum variation $t_1$, and hence no phase variation.
The period of \shb\ is found to be consistent with that of \citet{2015ApJ...799..154M}, 4.2$\pm$0.1 hr.
If we optimistically assume all wavelengths are statistically independent, the weighted average rotation period is 4.10 $\pm$ 0.03 hours.

Figure \ref{fig:amp2Spec} shows the posterior constraints on the secondary sinusoidal term.
As with $A_1$, the amplitude spectrum decreases as a function of wavelength, indicating Mie scattering grains and that the higher frequency term is likely just smaller scale cloud structures with similar properties as the longer timescale variability.
As with Figure \ref{fig:ampSpec}, we find no significant wavelength dependence on the time of maximum variation, $t_2$.

The same trend of decreasing variability amplitude for \shb\ was observed by HST's WFC3 grism over a shorter wavelength range (1.1 to 1.7~$\mu$m) by \citet{2015ApJ...798L..13Y}, shown in Figure \ref{fig:ampspec1821}.
The HST WFC3 maximum over minimum flux spectrum ($f_R$) is converted to an amplitude spectrum by assuming that the amplitude is half the difference in maximum and minimum flux, which can be shown to be 
\begin{equation}\label{eq:ampFromRatio}
A  = \frac{f_R - 1}{f_R + 1}
\end{equation}
There is a multiplicative offset of 25\% between the IRTF and WFC3 spectrum that may be due to different spatial cloud and temperature distributions at different epochs.
\citet{2015ApJ...798L..13Y} found that the WFC3 amplitude spectrum has no significant water vapor absorption, which indicates a high haze layer in {\shb}. This is consistent with our gradually decreasing slope from 1.1 to 1.7~$\mu$m.

\subsection{Mie Scattering Fit}

\begin{figure}
\begin{centering}
\includegraphics[width=0.5\textwidth]{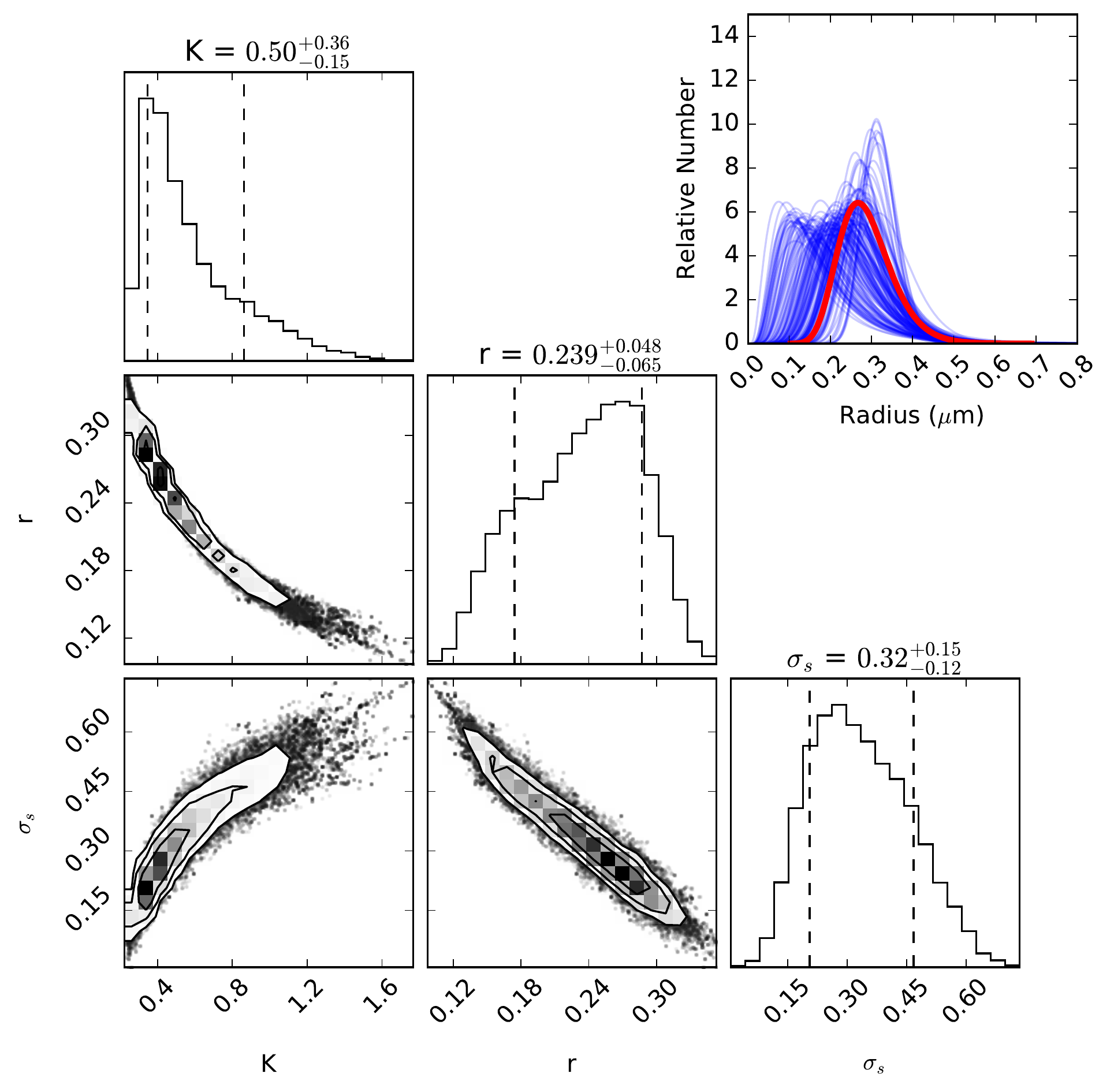}
\caption{Posterior distributions to the simple Mie scattering fit to the $A_1$ amplitude spectrum for \shb\ assuming a forsterite composition.
The model includes an amplitude offset $K$ to account for areal coverage, a median particle radius $r$ and a scale parameter $\sigma_s$.
The upper right inset plot shows the log-normal particle size distribution in blue for samples drawn from the posterior distribution.
The red curve is the maximum likelihood log-normal particle size distribution.}\label{fig:corner1821mie}
\end{centering}
\end{figure}

We model the wavelength dependence of variability amplitudes for \shb\ with a simple cloud model: an optically thin grouping of clouds consisting of a log-normal distribution of particle sizes.
Let $\alpha_{min}$ and $\alpha_{max}$ be the minimum and maximum cloud coverages corresponding to the maximum and minimum flux levels, respectively.
If we assume that the clouds are optically thin ($\tau \ll 1$), the ratio of maximum over minimum flux is
\begin{equation}
f_{R,Clouds} = \frac{1 - \tau \alpha_{min}}{1 - \tau \alpha_{max}},
\end{equation}
where $\tau$ is the clouds' optical depth.
Using Equation \ref{eq:ampFromRatio}, the amplitude is
\begin{equation}
A_{Clouds} = \frac{\tau \left(\alpha_{max} - \alpha_{min} \right)}{2 - \tau \left(\alpha_{max} + \alpha_{min} \right)}.
\end{equation}
In the optically thin case ($\tau \ll1$) the linear term dominates, so the amplitude, $A_{Clouds}$, is proportional to the optical depth $\tau$.
For the particle size distribution we assume a log-normal distribution with a median radius $r$ and width (scale parameter) $\sigma_s$ so our model reduces to
\begin{equation}
A_{Clouds} \approx K Q_{avg}(r,\sigma_s)
\end{equation}
where $A_{Clouds}$ is the variability amplitude, $K$ is a constant encapsulating the area and column density and $Q_{avg}(r,\sigma_s)$ is the extinction coefficient averaged over all particles.

We model the single particle extinction coefficient, $Q_{ext}$ with the \texttt{Python} Mie theory code \texttt{miescatter} \citep{bohren1983mie} to calculate the opacity as a function of wavelength.
We assume that the particles are spherical forsterite grains (Mg$_2$SiO$_4$).
We use a constant n=1.67 and k=0.006 complex index of refraction because the real and imaginary parts of the index of refraction are relatively constant across the optical and near infrared \citep{scott1996forsterite}.
When fitting the amplitude spectra to our Mie model, we imposed a prior of $r < 5 \mu m$; beyond this scale, the models are essentially unchanged with median particle radius.

For \shb, we find that the amplitude spectrum is well-fit by a Mie scattering model and that the median particle radius is reasonably constrained.
Figure \ref{fig:ampspec1821} shows the best-fit model, which has a median particle radius of 0.28~$\mu$m and $\sigma_s$ = 0.3.
This model has a reduced $\chi^2$ of 1.08 and is thus consistent with the measured amplitude spectrum.

Figure \ref{fig:corner1821mie} shows the posterior distribution for the $K$, $r$ and $\sigma_s$ parameters.
The derived median particle radius is $r$ = 0.24$^{+0.05}_{-0.07}$~$\mu$m 
and $\sigma_s$ = 0.32$^{+0.15}_{-0.12}$.
We show the maximum likelihood particle size distribution in the inset of Figure \ref{fig:corner1821mie} as well as a series of other possible distributions from the MCMC fitting.
Considering the large number of possible grain geometries and compositions, this is an underestimate for the true uncertainty in particle sizes.
Future work could include full modeling of the atmosphere with more complex spatial distributions, grain geometries and compositions (e.g., \citealt{2008ApJ...675L.105H}).

A similar amplitude versus wavelength trend was recently observed for WISEP J004701.06+680352.1 \citep{2016ApJ...829L..32L}, which has a redder than average ($J-K$ = 2.55) spectral energy distribution thought to be caused by a high altitude haze layer.
In WISEP J0047+6803, the wavelength trend could be fit by $\sim0.4~\mu$m forsterite grains, which is close to our characteristic particle size.

\citet{hiranaka2016subMicronDust} find that the observed spectra of reddened L dwarfs can be explained by extinction from a haze layer of sub-micron forsterite grains.
They find typical mean particle radii of 0.15 to 0.35 $\mu$m for a sample of L dwarfs.
Our median particle radius of $r$=0.24$^{+0.05}_{-0.07}$~$\mu$m for \shb\ is consistent with this range and may represent a partially cloudy version of the reddened brown dwarfs in \citet{hiranaka2016subMicronDust} or WISEP J0047+6803.

\section{Conclusions}\label{sec:conclusions}

We obtained high precision ground-based spectrophotometry of two brown dwarfs of spectral types L4-L5: {\sha} and {\shb}.
These brown dwarfs have temperatures near the condensation temperature of forsterite (Mg$_2$SiO$_4$).
We find that {\sha} has marginal variability and no significant wavelength dependence; while
{\shb} shows significant variability up to the 1.5\% amplitude (3.0\% peak-to-valley) at short wavelengths (0.9~$\mu$m), which declines toward the near infrared (2.4$~\mu$m).
A possible difference between the two brown dwarfs is the viewing geometry with \sha\ closer to pole-on and \shb\ closer to equator-on as shown in Table \ref{tab:bdProp}.
Recently, \citet{vos2017viewingGeom} find that there are correlations between variability and color properties as a function of viewing geometry and that equator-on dwarfs show, on average, larger variability amplitudes.
\sha\ shows less variability than reported in the literature \citep{2004MNRAS.354..378K,2014A&A...566A.111W}, so it likely exhibited a more homogeneous cloud distribution at our observed epoch.

The variability of {\shb} is consistent with previous HST measurements for the system that also show a gradual decrease in variability amplitude with wavelength from 1.1~$\mu$m to 1.7 $\mu$m with no decrease in the water vapor band \citep{2016ApJ...826....8Y}.
The lack of the variability in the water vapor bands in both the WFC3 data and the data presented here supports evidence of a high altitude aerosol layer in the atmosphere of this L dwarf.
We find no statistically significant phase offsets in the light curves for \sha\ nor \shb\ from wavelength to wavelength.
This is different from  \citep{2016ApJ...826....8Y} who found a significant phase offset using simultaneous {\em Spitzer} and {\em HST} light curves at two different epochs.
This may be due to the larger differences in pressures probed by the {\em Spitzer} observations (3.6 $\mu$m and 4.5 $\mu$m), not sampled by our more restricted wavelength range.

We fit the variability amplitude spectrum of {\shb} with an optically thin Mie scattering model and find it is consistent with forsterite grains with sub-micron sized particles.
A log-normal particle size distribution reproduces the data with a median grain radius of 0.24$^{+0.05}_{-0.07}$ $\mu$m.
However, considering different particle compositions and grain geometries can widen the allowed range of particle sizes.

Using ground-based facilities such as SpeX on IRTF allows us to expand our monitoring of cloud-bearing ultracool dwarfs, both in sample size and long-term tracking. The former can provide powerful constraints on how dust particle sizes may depend on differing pressure-temperature profiles, rotation rates, composition and viewing orientations while the latter allows us to study weather-related changes to cloud structures over multiple rotation periods. We are currently pursuing a larger monitoring sample that will be reported in future publications.

\section{Acknowledgements}
Thanks to the anonymous referee for useful improvements and corrections to this work.
Thanks to Vivien Parmentier for useful discussions on cloud versus a spotted temperature models.
Funding for the E Schlawin is provided by NASA Goddard Spaceflight Center.
The material is based upon work supported by
the National Aeronautics and Space Administration under
Grant No.~NNX15AI75G. 
This research has made
use of the SIMBAD database, operated at CDS, Strasbourg,
France; NASA's Astrophysics Data System Bibliographic
Services; the M, L, T, and Y dwarf compendium
housed at \url{http://DwarfArchives.org}; the SpeX Prism
Libraries at \url{http://www.browndwarfs.org/spexprism}.
The authors wish to recognize and acknowledge the very significant cultural role and reverence that the summit of Mauna Kea has always had within the indigenous Hawaiian community. We are most fortunate to have the opportunity to conduct observations from this mountain.

\acknowledgments

\facility{IRTF (SpeX)}.

\software{
\texttt{emcee} \citep{foreman-mackey2013emcee},
\texttt{corner.py} \citep{foremanCorner},
\texttt{miescatter} \citep{bohren1983mie},
SpeX Prism Library Analysis Toolkit \citep{2017arXiv170700062B},
\texttt{astropy} \citep{astropy2013},
\texttt{IRAF} \citep{tody86} and
\texttt{IDL} version 8.5 (Exelis Visual Information Solutions, Boulder, Colorado)}

\appendix

\section{Original Light Curves}

\begin{figure*}[!t]
\centering
\subfloat[2MASS J0835]{
	\includegraphics[width=0.5\textwidth]{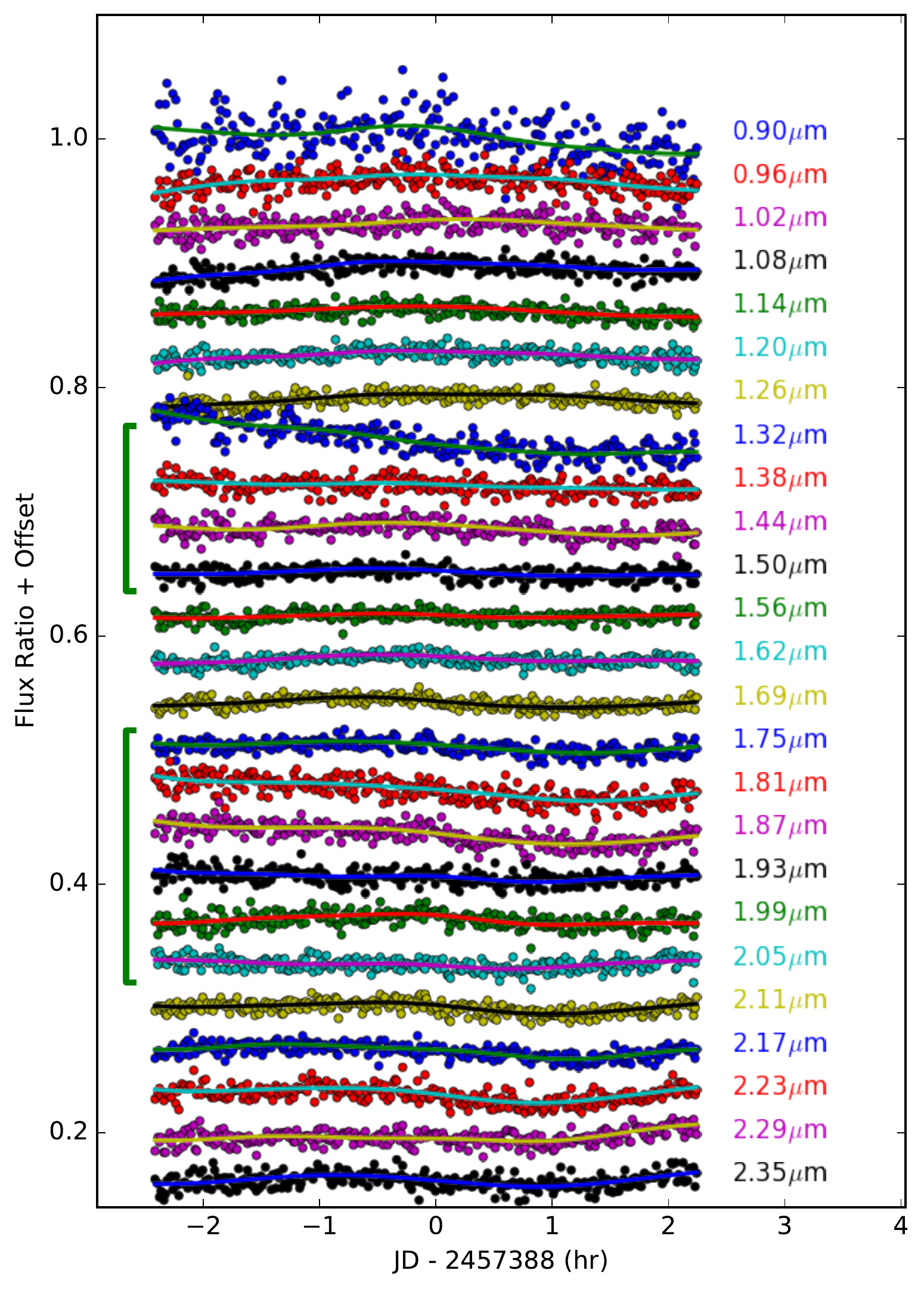}
	\label{fig:tserPhase0835NoDetrend}
	}
\subfloat[2MASS J1821]{
	\includegraphics[width=0.5\textwidth]{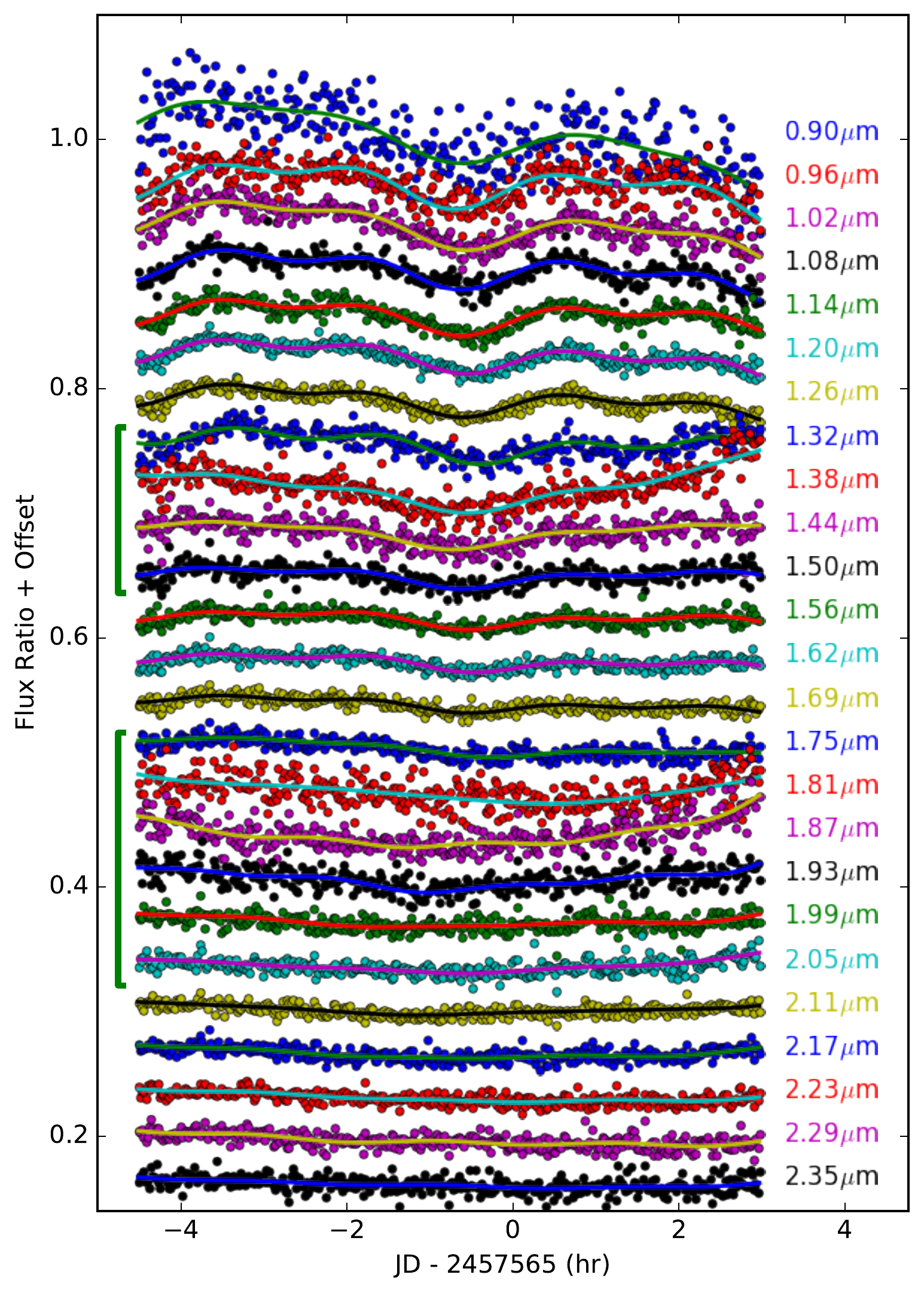}
	\label{fig:tserPhase1821NoDetrend}
	}
	\caption{Time series for each system with an offset added for clarity, as in Fig \ref{fig:tserDetrend}, but in this case showing the raw light curves without de-trending.
	Each wavelength bin is labeled on the right side. Significant variations in the baseline are modeled as linear functions of airmass and time which are most prominent near edges of telluric absorption features at 1.3$\mu$m and 1.8$\mu$m.
	Green brackets show regions where telluric absorption can contaminate the time series.}
	\label{fig:tserNoDetrend}
\end{figure*}

We show the original light curves used in model fitting in Figure \ref{fig:tserNoDetrend}, which preserve the significant variations of flux as a function of time and airmass.
Our sinusoidal model using Equation \ref{eq:cosfit} is shown for each of the light curves in Figure \ref{fig:tserNoDetrend}.
For Figure~\ref{fig:tserDetrend} we remove the baseline variations to show the astrophysical modulations due to brown dwarf rotation.

\bibliographystyle{apj}
\bibliography{bd_biblio}

\end{document}